\RequirePackage{fix-cm}
\documentclass{svjour3}                     
\smartqed  
\usepackage{graphicx}
\usepackage{amssymb}
\usepackage{amsmath}
\usepackage{algorithm2e}
\usepackage{bm}

\journalname{Designs, Codes and Cryptography}
\begin{document}

\title{On the bounds and achievability about the ODPC of $\mathcal{GRM}(2,m)^*$ over prime field
for increasing message length
\thanks{This work is supported in part by the National Key Basic Research and Development Plan of China under Grant 2012CB316100, and
the National Natural Science Foundation of China under Grants 61271222, 60972033.}
}

\titlerunning{Bounds and achievability about the ODPC of $\mathcal{GRM}(2,m)^*$}

\author{Xiaogang Liu \and Yuan Luo}


\institute{Xiaogang Liu
\at Department of Computer Science and Engineering, Shanghai Jiao Tong University, China
 \\\email{liuxg0201@163.com}
\and
Yuan Luo (Corresponding author)
\at Department of Computer Science and Engineering, Shanghai Jiao Tong University, China
\\Addr.: 800 Dongchuan Road, Min Hang District, Shanghai 200240, China.
\\\email{yuanluo@sjtu.edu.cn}
}

\date{Received: date / Accepted: date}
\maketitle

\begin{abstract}
  The optimum distance profiles of linear block codes were studied for increasing or decreasing message length while keeping the minimum distances as large as possible, especially for Golay codes and the second-order Reed-Muller codes, etc. Cyclic codes have more efficient encoding and decoding algorithms. In this paper, we investigate the optimum distance profiles with respect to the cyclic subcode chains (ODPCs) of the punctured generalized second-order Reed-Muller codes $\mathcal{GRM}(2,m)^*$ which were applied in Power Control in OFDM Modulations in channels with synchronization, and so on.
For this, two standards are considered in the inverse dictionary order, i.e., for increasing message length.
Four lower bounds and upper bounds on ODPC are presented, where the lower bounds almost achieve the corresponding upper bounds in some sense. The discussions are over nonbinary prime field.

\end{abstract}

\section{Introduction} \label{Sec1}

In some communication systems, the number of transmitted information bits varies according to the situations, such as in designing the transport format combination indicators in CDMA systems when the size of users increases or decreases, see \cite{HT001,TJ001}. For those cases, the distance profiles of the linear block codes were introduced and designed to be optimum to keep the minimum distance as large as possible in the variation process  \cite{CH001,HL001,HL002,MSL001}.

This subject was introduced in \cite{HL001}, and then studied in \cite{CH001,HL002} for linear block codes, i.e., the generalized Reed-Solomon codes, the Golay code and its extension, the first-order Reed-Muller codes and the second-order Reed-Muller codes, etc. Because of the convenience for encoding and decoding, the optimum distance profile with respect to the cyclic subcode chain (ODPC) of cyclic code was presented in \cite{LH003}, and investigated for the punctured second-order Reed-Muller codes $\mathcal{RM}(2,m)^*$ in \cite{LLK002}. For a general cyclic code, there may not be a routine method for determining the ODPC.
 In this paper, we focus on the punctured generalized second-order Reed-Muller codes $\mathcal{GRM}(2,m)^*$ and a class of cyclic subcodes.

As a geometrical code, Reed-Muller code can be decoded by majority logic decoding, list decoding \cite{E001,W001} and soft-decision decoding \cite{BD001}.
It can be used for locating malicious nodes \cite{KW001}, and constructing polar codes \cite{A002,SE001}.
 Generalized Reed-Muller codes received more and more attention \cite{B001,G001}, one important application of which was in Power Control in OFDM Modulations \cite{DJ002,KGP001}.

For the punctured second-order Reed-Muller codes $\mathcal{RM}(2,m)^*$, some ODPCs were presented for the case when $m$ is even, and some suboptimum results were given when $m$ is odd \cite{LLK002}.
Considering the increasement or decreasement of the message length with respect to the required dimension or not, the ODPCs are investigated in the inverse dictionary order and dictionary order under Standards I or II, respectively.
The frame of this paper focuses on these problems about $\mathcal{GRM}(2,m)^*$ with increasing message length.

 In Section \ref{Sec2}, basic definitions are provided around the ODPC.
 Section \ref{Sec3} explains the basic notations about the generalized Reed-Muller codes $\mathcal{GRM}(\mu,m)$,
 quadratic forms and exponential sums. Then the ODPCs-II$^{inv}$ of a class of cyclic subcodes are obtained for most cases of $m$.
More relations among alternating bilinear forms and quadratic forms are presented in Section \ref{Sec4}, from which the ODPCs of $\mathcal{GRM}(2,m)^*$ are investigated in the inverse dictionary order under two standards.
 In fact, an upper bound is given which can be almost achieved by our lower bound, refer to Corollary \ref{CDP001}.
Section \ref{Sec5} is a final conclusion.
{\bf Note that, here $\mathcal{GRM}(2,m)^*$ is over an odd prime field $\mathbb{F}_p$}.

\section{Preliminaries} \label{Sec2}

In this section, a brief explanation of the following definitions are presented:
distance profile of a linear block code (DPB),
 the optimum distance profile of a linear block code (ODPB),
 distance profile with respect to cyclic subcode chain of a cyclic code (DPC),
 and the optimum DPCs under two respective standards (ODPC-I and ODPC-II), etc.

\subsection{Optimum distance profiles and subcode chains of a linear block code\label{Sec2.1}}

Let $C$ be an $[n, k]$ linear code over $\mathbb{F}_q$ and denote $C_0=C$.  A sequence of linear subcodes
$C_0\supset C_1\supset\cdots\supset C_{k-1}$
is called a {\bf subcode chain}, where $\dim[C_i]=k-i$.
An increasing sequence
$d[C_0]\le d[C_1]\le\cdots\le d[C_{k-1}] $
is called a {\bf distance profile of the linear block code $C$ (DPB)},
where $d[C_i]$ is the minimum Hamming distance of the subcode $C_i$ \cite{HL001}.

We say that integer sequence $a_0, \ldots, a_{k-1}$ is
larger than $b_0, \ldots, b_{k-1}$ in the inverse dictionary order if
there is an integer $t$ such that $a_t > b_t$ and $a_i= b_i$ for $k-1\ge i\ge t+1$.
  $a_0, \ldots, a_{k-1}$ is an upper bound on
$b_0, \ldots, b_{k-1}$ in that order if
$a_0, \ldots, a_{k-1}$ is larger than or equal to $b_0, \ldots, b_{k-1}$.
The optimum distance profile of $C$ denoted by ODPB$[C]^{inv}$:
\begin{equation}\label{eq10}
ODPB[C]^{inv}_0, ODPB[C]^{inv}_1, \cdots, ODPB[C]^{inv}_{k-1},
\end{equation}
  is an upper bound on any distance profile in that order.
 A subcode chain that achieves the optimum distance profile is called an {\bf optimum chain} \cite{HL001}.
The inverse dictionary order corresponds to the problem of increasing the message length.

\subsection{Distance profiles with respect to cyclic subcode chains\label{Sec2.3}}

For a cyclic $[n, k]$ code ${C}$ over $\mathbb{F}_q$, where $\mbox{gcd}(n, q)=1$,
{\bf a cyclic subcode chain} of ${C}$ is a chain of cyclic subcodes such that:
 ${C}_{\tau_0}\supset {C}_{\tau_1}\supset\cdots\supset {C}_{\tau_{\lambda-1}}\supset \{0^n\}$,
where ${C}_{\tau_0}={C}$ and there is no cyclic subcode between any two neighbors in the chain.
The increasing sequence
\[
d[{C}_{\tau_0}]\le d[{C}_{\tau_1}]\le\cdots\le d[{C}_{\tau_{\lambda-1}}]
\]
is called the {\bf distance profile with respect to the cyclic subcode chain (DPC)},
where $\lambda$ is called the length of the profile or the length of the chain \cite{LH003}. The decreasing sequence
\[
\dim[{C}_{\tau_0}]>\dim[{C}_{\tau_1}]>\dim[{C}_{\tau_2}] > \cdots > \dim[{C}_{\tau_{\lambda-1}}],
\]
is called the {\bf dimension profile with respect to the cyclic subcode chain}. In general, ${C}_{\tau_u}$ denotes a cyclic subcode in a chain, and math calligraphy $\mathcal{C}_i$ denotes an irreducible cyclic code.

In the comparison among the DPCs in the inverse dictionary order,
according to the dimension profiles or not, two standards are introduced as follows respectively.
For a given cyclic code ${C}$, the lengths of its DPCs are the same, see \cite{LH003}.
Two chains with length $\lambda$ are set to be in the same class if
$\dim[{C}^1_{\tau_u}]=\dim[{C}^2_{\tau_u}]\, \ \mbox{for $0\le u\le \lambda-1$},$
where the superscripts 1 and 2 denote the two chains respectively.
 The analysis under {\bf Standard I} is to find the optimum DPC denoted by {\bf ODPC-I$^{inv}$} for each class in the inverse dictionary order. This idea is with respect to the variation of the transmission rate, or equivalently the dimension profile.
 Some counting properties of the classification are presented in Section \ref{Sec2.4}.

The analysis without the dimension profile is said to be under {\bf Standard II}.
The distance profiles of any two chains are compared  directly in the inverse dictionary order to obtain the optimum one, which is denoted by {\bf ODPC-II$^{inv}$}. That is to say the minimum distance receives more attention.

A cyclic subcode chain that achieves the ODPC (I or II) is called an {\bf optimum cyclic subcode chain} correspondingly, and
the optimum one among all the ODPC-Is of the classes is the ODPC-II.

\subsection{Key parameters of cyclic subcode chains} \label{Sec2.4}

For a cyclic code with generator polynomial $g(x)$, let $P$ be the set of
minimal polynomials that are factors of $g(x)$, and $J(v)$ be the number of polynomials with degree $v$ in $P$.

\begin{lemma}\label{th1} (Theorem 1, \cite{LH003})
  Let ${C}$ be an $[n, k]$ cyclic code over $\mathbb{F}_q$, and $m$ be the multiplicative order of $q$ modulo $n$, i.e. $\mbox{ord} (q, n)$.
\begin{itemize}
\item
The length of its cyclic subcode chains is $ \lambda=|A\setminus P|=\sum_{v: v|m} (L(v)-J(v))$,
where $L(v)$ is the number of $q-$cyclotomic cosets modulo $n$ with size $v$, i.e. $ L(v)=\sum_{g\in G(v)} \frac{\varphi(n/g)}{v}$,
$G(v)=\{g: v=\mbox{ord}(q, n/g), g|n\}$ and $\varphi(\cdot)$ is the Euler function.
\item
The number of its cyclic subcode chains is
$\lambda!,$
i.e. $\lambda$ factorial.
\item
The number of the chains in each class is $\mu=\prod_{v: v|m} (L(v)-J(v))!$.
\item
The number of the classes is
$\frac{\lambda!}{\mu}.$
\end{itemize}
Note that, the integers modulo $n$ are considered in $\{1, 2,\cdots, n\}$.
\end{lemma}
For examples, refer to \cite{LLK002,LH003}.

\section{The ODPCs of a class of cyclic subcodes of $\mathcal{GRM}(2,m)^*$}\label{Sec3}

There are four subsections in this section. Section \ref{Sec3.1} explains the basic definitions and notations about the generalized Reed-Muller codes $\mathcal{GRM}(\mu,m)$.
Some relevant knowledge about quadratic forms and exponential sums is presented in Section \ref{Sec3.2}. The weights and weight distributions of some cyclic codes with unique primitive idempotent are analyzed in Section \ref{Sec3.3}, which induce Theorem \ref{F01} in Section \ref{Sec3.5} where the ODPCs-II$^{inv}$ of a class of cyclic subcodes are obtained for most cases of $m$.

\subsection{The generalized Reed-Muller codes $\mathcal{GRM}(\mu,m)$}\label{Sec3.1}

 In this subsection, we give an elementary account of the generalized Reed-Muller codes, including the basic definitions and important properties of the punctured generalized Reed-Muller codes, such as the dimensions and generator polynomials. Especially, Lemma \ref{R002} implies that $\mathcal{GRM}(\mu,m)^*$ is a cyclic code with the construction of idempotent (\ref{PI001}), which supports the generation of the cyclic subcode chains.
 More detailed work and descriptions can be found in \cite{DG001,KLP001,W01}.

The general definition is based on an arbitrary finite field $\mathbb{F}_q$ with $q$ elements where $q$ is a power of an odd prime $p$. We consider the generalized Reed-Muller codes over the field $\mathbb{F}_p$.

     Take $V$ to be the space $\mathbb{F}_p^m$ of $m$-tuples, with standard basis $\bf{e_1},\ldots,\bf{e_m}$, where $\bf{e_i}$$ =(0,0,\ldots,0,1,0,\ldots,0)$ (with 1 in the $i$th position)
and a general vector in $V$ is denoted by $v$.
The generalized Reed-Muller codes are $p$-ary codes, and the ambient space will be the function space $\mathbb{F}_p^V$, with the usual basis of characteristic functions of the vectors of $V$. Denote $f\in \mathbb{F}_p^V$ by $
f=f(x_1,\ldots,x_m)$ where $x_i$ lies in $\mathbb{F}_p$. Set $\mathcal{M}$ to be the set of $p^m$ monomial polynomials
\[
\mathcal{M}=\{x_1^{i_1}x_2^{i_2}\cdots x_m^{i_m}|0\leq i_k\leq p-1, k=1,2,\ldots,m\}.
\]
Then $\mathcal{M}$ forms another basis of $\mathbb{F}_p^V$.
  The following two definitions and three lemmas are well-known results \cite{AK001}.

\begin{definition}
For prime field $\mathbb{F}_p$, set $V=\mathbb{F}_p^m$. Then for any $\mu$ such that $0\leq \mu \leq m(p-1)$, the {$\bm{\mu}^{\mbox{th}}$} {\bf order generalized Reed-Muller codes $\mathcal{GRM}(\mu,m)$} over $\mathbb{F}_p$ is the subspace of $\mathbb{F}_p^V$ (with basis of characteristic functions on the vectors in $V$) of all reduced $m$-variable polynomial functions of degree at most $\mu$. Thus
\[
\mathcal{GRM}(\mu,m)=\left<x_1^{i_1}x_2^{i_2}\cdots x_m^{i_m}|\sum_{k=1}^{m}i_k \leq \mu \right>.
\]
\end{definition}

\begin{lemma}
For any $\mu$ such that $0\leq \mu \leq m(p-1)$,
\[
\operatorname{dim}\left(\mathcal{GRM}(\mu,m)\right)=\sum_{i=0}^{\mu}\sum_{k=0}^{m}(-1)^k\left(\begin{array}{c}m\\k\end{array}\right)
\left(\begin{array}{c}i-kp+m-1\\i-kp\end{array}\right).
\]
Especially, $\mbox{dim}\left(\mathcal{GRM}(2,m)\right)=1+m+{{m(m+1)}\over 2}$. Here, $\left(\begin{array}{c}m\\k\end{array}\right)$ represents the general binomial coefficient.
\end{lemma}

The {$\bm{\mu}^{\mbox{th}}$ } {\bf order punctured generalized Reed-Muller codes}, where $0\leq \mu \leq m(p-1)$, denoted by $\mathcal{GRM}(\mu,m)^*$, is the cyclic code of length $p^m-1$ obtained by deleting the coordinate position ${\bf 0}$ from $\mathcal{GRM}(\mu,m)$.

\begin{lemma}\label{MRM01}
If $\mu=r(p-1)+s<m(p-1)$ with $0\leq s<p-1$, then $\mathcal{GRM}(\mu,m)^*$ has minimum weight $(p-s)p^{m-r-1}-1$.
\end{lemma}

\begin{definition}\label{R02}
For $p$ a prime and $u$ a positive integer with $p$-ary representation $u=\sum_{i=0}^{\infty}u_ip^i, \ \ \mbox{where} \ \ 0\leq u_i\leq p-1$,
the $\bm{p}${\bf-weight} of $u$ is $\mbox{w}_p(u)$ given by $\mbox{w}_p(u)=\sum_{i=0}^{\infty}u_i$.
\end{definition}
For $p=2$, $2$-weight was used by MacWilliams to study the zeros of the punctured Reed-Muller codes \cite[pp.383]{MS001}. By using Definition \ref{R02}, Lemma \ref{R002} describes the zeros of $\mathcal{GRM}(\mu,m)^*$ with $p$-weight.

\begin{lemma}\label{R002}
Let $\pi$ be a primitive root of $\mathbb{F}_q$ where $q=p^m$, then, for $0\leq u\leq p^m-2$, $\pi^u$ is a root of the generator polynomial of the code $\mathcal{GRM}(\mu,m)^*$ if and only if $0<\mbox{w}_p(u)\leq m(p-1)-1-\mu$.
\end{lemma}

From Lemma \ref{R002}, it can be verified that $\mathcal{GRM}(2,m)^*$ is a cyclic code with idempotent
\begin{equation}\label{PI001}
\theta_0+\theta_1^*+\theta_{l_{\lfloor{m\over 2}\rfloor}}^*+\cdots+\theta_{l_1}^*+\theta_{l_0}^*
\end{equation}
where $l_i=1+p^i (0\leq i\leq {\lfloor{m/2}\rfloor})$.
In the subsequent sections, some cyclic subcodes of $\mathcal{GRM}(\mu,m)^*$ will be studied for the ODPCs. See Section \ref{Sec3.3} for the definition of primitive idempotent.

\subsection{Relevant results from finite fields} \label{Sec3.2}
In this subsection, relevant knowledge from finite fields is presented first for our study of cyclic codes in Section \ref{Sec3.2.1} \cite{LH001}. Then some results about the ranks of particular quadratic forms and the calculations of certain exponential sums are given in Section \ref{Sec3.2.2} and Section \ref{Sec3.2.3} respectively. Note that Lemma \ref{FL002} and Lemma \ref{DH001} are known results.

\subsubsection{Finite fields and cyclic codes} \label{Sec3.2.1}

 There is a lot of research on cyclic codes, see \cite{D01,HKM001,M001} for the irreducible case, and \cite{DLMZ001,FL001,LTW001} for the reducible case.
Here, some known properties are listed about the codeword weight, and the mathematical tools quadratic form and symmetric matrix. \\

\noindent{\sl Weight of codeword in cyclic codes:}

  Let the cyclic code ${C}$ over $\mathbb{F}_p$ be of length $l=q-1=p^m-1$ with  parity check polynomial
\begin{equation}
h(x)=h_1(x)\cdots h_{\iota}(x) \ \ \ (\iota \geq 1),
\end{equation}
where $h_\lambda(x) (1\leq \lambda \leq \iota)$ are distinct irreducible polynomials in $\mathbb{F}_p[x]$ with degrees $e_\lambda (1\leq \lambda\leq \iota)$, then $k=\mbox{dim}_{\mathbb{F}_p}{C} = \sum_{\lambda=1}^{\iota}e_\lambda \leq m\iota $. Let $\pi$ be a primitive element of $\mathbb{F}_q$ and $\pi^{-s_\lambda}$ be a zero of $h_\lambda(x), 1\leq s_\lambda \leq q-2 (1\leq \lambda \leq \iota)$. Then the codewords in ${C}$ can be expressed by
\begin{equation}\label{CW001}
c(\alpha_1,\ldots,\alpha_{\iota}) = (c_0,c_1,\ldots,c_{l-1}) \ \ (\alpha_1,\ldots,\alpha_{\iota}\in \mathbb{F}_q),
\end{equation}
where $c_i = \sum_{\lambda =1}^{\iota}\mbox{Tr}(\alpha_{\lambda}\pi^{is_{\lambda}}) (0\leq i\leq l-1)$ and $\mbox{Tr}:\mathbb{F}_q\rightarrow \mathbb{F}_p$ is the trace mapping from $\mathbb{F}_q$ to $\mathbb{F}_p$. Therefore the Hamming weight of the codeword $c=c(\alpha_1,\ldots,\alpha_{\iota})$ is:

\begin{equation}\label{C01}
\begin{array}{ll}
w_H(c)&=\# \{i|0\leq i \leq l-1, c_i\not= 0\}\\
      &=l-{l\over p}-{1\over p}\sum_{a=1}^{p-1}\sum_{x\in \mathbb{F}_q^*}{\zeta}_p^{\mbox{Tr}(af(x))}\\
      &=p^{m-1}(p-1)-{1\over p}\sum_{a=1}^{p-1}S(a\alpha_1,\ldots,a\alpha_{\iota})\\
      &=p^{m-1}(p-1)-{1\over p}R(\alpha_1,\ldots,\alpha_{\iota})
\end{array}
\end{equation}
where  ${\zeta_p}=e^{{2\pi i}\over p}$ ($i$ is the imaginary unit), $f(x)=\alpha_1x^{s_1}+\alpha_2x^{s_2}+\cdots +\alpha_{\iota}x^{s_{\iota}}\in \mathbb{F}_q[x], \mathbb{F}_q^*=\mathbb{F}_q\backslash \{0\}$,
\begin{equation}\label{ES01}
S(\alpha_1,\ldots,\alpha_{\iota}) = \sum_{x\in \mathbb{F}_q}{\zeta}_p^{\mbox{Tr}\left(\alpha_1x^{s_1}+\cdots +\alpha_{\iota}x^{s_{\iota}}\right)},
\end{equation}
and $R(\alpha_1,\ldots,\alpha_{\iota})=\sum_{a=1}^{p-1}S(a\alpha_1,\ldots,a\alpha_{\iota})$.

\begin{remark}
There may not be a one-to-one correspondence between the codewords of ${C}$ and equation (\ref{CW001}).
\end{remark}

\begin{remark}
When the primitive idempotent $\theta_0$ ($s_i=0$ for one $i$) is included in the construction of the idempotent of corresponding cyclic code, equation (\ref{C01}) changes to
\begin{equation}\label{C02}
w_H(c) =p^{m-1}(p-1)+\delta_{b,0}-1-{1\over p}\sum_{a=1}^{p-1}{\zeta_p^{ab}}S(a\alpha_1,\ldots,a\alpha_{\iota})
\end{equation}
for some $b\in \mathbb{F}_p$, where $\delta_{b,0}$ is $1$ for $b=0$, and $0$ otherwise.
\end{remark}

\noindent {\sl Quadratic forms:}

 Fix a basis $v_1,\ldots,v_m$ of $\mathbb{F}_q$ over $\mathbb{F}_p$ where $q=p^m$, then each $x\in \mathbb{F}_q$ can be uniquely expressed as
 $x=x_1v_1+\ldots +x_mv_m \  (x_i \in \mathbb{F}_p)$,
which is used in the following $\mathbb{F}_p$-linear isomorphism: $\mathbb{F}_q \overset{\sim}{\rightarrow} \mathbb{F}_p^m, \ x \mapsto X=(x_1,\ldots,x_m)$.
Therefore, a function $f:\mathbb{F}_q \rightarrow \mathbb{F}_q$ induces a function $F:\mathbb{F}_p^m \rightarrow \mathbb{F}_p$ where $F(X)=\mbox{Tr}(f(x))$. For general functions of the form
\begin{equation}\label{QF03}
f_{\alpha,\ldots,\beta}(x) = \alpha x^{p^i+1} + \cdots + \beta x^{p^j+1}
\end{equation}
 where $0\leq i,\ldots, j \leq  \lfloor{{m}\over 2}\rfloor$, there exist quadratic forms
 \begin{equation}\label{QF01}
 F_{\alpha,\ldots,\beta}(X)
 \end{equation}
  and corresponding symmetric matrices $H_{\alpha,\ldots,\beta}$
  satisfying $F_{\alpha,\ldots,\beta}(X)=XH_{\alpha,\ldots,\beta}X^T=\mbox{Tr}(f_{\alpha,\ldots,\beta}(x)).$

  \hspace{0.5cm}

\noindent{\sl Symmetric matrices:}

For an ${m\times m}$ symmetric matrix $H$ over $F_p$ and $r= \mbox{rank}H$, there exists $M \in \mbox{GL}_m(\mathbb{F}_p)$ such that $H'=MHM^T=\mbox{diag}(a_1,\ldots,a_r,0,\ldots,0)$
where $a_i\in \mathbb{F}_p^*$. Let $\Delta =a_1\cdots a_r$ (set $\Delta =1$ for $r=0$), and $\left({\Delta \over p}\right)$
denotes the {\bf Legendre symbol}.
We have the following result about the exponential sum corresponding to the matrix $H$.

\begin{lemma}(Lemma 1, \cite{FL001})\label{FL002}
\begin{enumerate}
\renewcommand{\labelenumi}{$($\mbox{\roman{enumi}}$)$}
\item
For the quadratic form $F(X) = XHX^T$, 
\begin{equation}\label{QF02}
\sum_{X\in \mathbb{F}_p^m}\zeta_p^{F(X)} =
\begin{cases}
\left({\Delta \over p}\right)p^{m-r/2} & \mbox{if} \ \ p\equiv 1 \ (\mbox{mod} \ 4),\\
i^r\left({\Delta \over p}\right)p^{m-r/2} & \mbox{if} \ \ p \equiv 3 \ (\mbox{mod} \ 4).
\end{cases}
\end{equation}
\item
For $A = (a_1,\ldots,a_m)\in \mathbb{F}_p^m$, if \ $2YH+A=0$ has solution $Y=B \in \mathbb{F}_p^m$, then
\begin{equation}\label{EC01}
\sum_{X\in \mathbb{F}_p^m}{\zeta}_p^{F(X)+AX^T} = {\zeta}_{p}^c\sum_{X\in \mathbb{F}_p^m}{\zeta}_p^{F(X)} \ \mbox{where} \ \ c={1\over 2}AB^T\in \mathbb{F}_p.
\end{equation}
Otherwise $\sum_{X\in \mathbb{F}_p^m}{\zeta}_p^{F(X)+AX^T} = 0$.
\end{enumerate}
\end{lemma}

\subsubsection{The ranks of particular quadratic forms}\label{Sec3.2.2}

 In this subsection, the ranks of some quadratic forms corresponding to particular functions are studied, where Lemma \ref{MS02} concerns the case for multiple items and Lemma \ref{R001} concerns the case with only one item.

For $m$ a positive integer, set
\[
f_d(x) = {\gamma_{\lfloor {m\over 2}\rfloor}x^{p^{\lfloor {m\over 2}\rfloor}+1}} + {{\gamma_{\lfloor {m\over 2}\rfloor-1}x^{p^{\lfloor {m\over 2}\rfloor-1}+1}}+\cdots+{\gamma_d}x^{p^d+1}},
\]
 which induces a quadratic form
\[
 F_d(X) = \mbox{Tr}(f_d(x))= \mbox{Tr}(f_d(x_1v_1+\cdots+x_mv_m)) = XH_dX^T
\]
  where $H_d$ is the corresponding symmetric matrix. Denote $r_d = \mbox{rank}(H_d)$.

Combining the methods used in \cite{CH001} and \cite{FL001}, there is the following result.
\begin{lemma} \label{MS02}
Let $m\geq 3$ be a positive integer and $0\leq d\leq {\lfloor {m\over 2}\rfloor}$. For $(\gamma_{\lfloor {m\over 2}\rfloor},\gamma_{\lfloor {m\over 2}\rfloor-1},\ldots,\gamma_d)\in \mathbb{F}_q^{{\lfloor {m\over 2}\rfloor}-d+1}\backslash\{(0,0,\ldots,0)\}$, $r_d\geq 2d$ if $H_d\not= 0_{m\times m}$.
\end{lemma}

\begin{proof}
Only the case of $m=2t+2$ is considered here, which implies that $\lfloor {m\over 2}\rfloor=t+1$.

For $Y = (y_1,\ldots,y_m)\in \mathbb{F}_p^m$ and $y = y_1v_1+\cdots+y_mv_m\in \mathbb{F}_q$,
\begin{equation}\label{LM004}
F_d(X+Y) - F_d(X) -F_d(Y) = 2YH_dX^T
\end{equation}
and
\begin{equation}\label{LM001}
\begin{array}{l}
\mbox{Tr}({f_d(x+y)})-\mbox{Tr}({f_d(x)})-\mbox{Tr}({f_d(y)}) \\
 = \mbox{Tr}\left(\left(\gamma_{t+1}x^{p^{t+1}}y+\gamma_{t+1}y^{p^{t+1}}x\right)  +  \left(\gamma_{t}x^{p^{t}}y  + \gamma_{t}y^{p^{t}}x \right) + \cdots + \left(\gamma_{d}x^{p^{d}}y + \gamma_{d}y^{p^{d}}x \right)\right)  \\
 = \mbox{Tr}\left(y  \left(\left(\gamma_{t+1}x^{p^{t+1}}+\gamma_{t+1}^{p^{t+1}}x^{p^{t+1}} \right) +  \left( \gamma_{t}x^{p^{t}}+\gamma_{t}^{p^{t+2}}x^{p^{t+2}}\right)  + \cdots+ \left(\gamma_{d}x^{p^{d}}+\gamma_{d}^{p^{m-d}}x^{p^{m-d}}\right) \right)\right)\\
 =\mbox{Tr}\left(y\phi_d(x)\right)
\end{array}
\end{equation}
where we have used the fact that $\mbox{Tr}(\alpha + \beta) = \mbox{Tr}(\alpha) + \mbox{Tr}(\beta)$ and $\mbox{Tr}(\alpha) = \mbox{Tr}(\alpha^p)$ for $\alpha, \beta \in F_q$. Here
\begin{equation}\label{LM002}
\phi_d(x) = \gamma_{t+1}x^{p^{t+1}}+\gamma_{t+1}^{p^{t+1}}x^{p^{t+1}}  +  \gamma_{t}x^{p^{t}}+\gamma_{t}^{p^{t+2}}x^{p^{t+2}}  + \cdots+ \gamma_{d}x^{p^{d}}+\gamma_{d}^{p^{m-d}}x^{p^{m-d}}
 \end{equation}
 which is a $p$-polynomial over $\mathbb{F}_q$. 
 Since $0\leq d\leq t+1$, we have $m-d\geq t+1\geq d$, $\mbox{deg}(\phi_d(x))\leq p^{m-d}$ and the smallest degree of $x$ in $\phi_d(x)$ is not less than $p^d$. So,
\begin{equation}\label{LM003}
\begin{array}{l}
\phi_d(x) \\
= \left( \gamma_{t+1}^{p^{m-d}}x^{p^{t+1-d}}+\gamma_{t+1}^{p^{t+1-d}}x^{p^{t+1-d}}  +  \gamma_{t}^{p^{m-d}}x^{p^{t-d}}+\gamma_{t}^{p^{t+2-d}}x^{p^{t+2-d}}  + \cdots+\gamma_{d}^{p^{m-d}}x+\gamma_{d}^{p^{m-2d}}x^{p^{m-2d}}\right)^{p^d}\\
=\phi_d'(x)^{p^d}
\end{array}
\end{equation}
where $\mbox{deg}(\phi_d'(x)) \leq p^{m-2d}$. Therefore by using (\ref{LM004}), (\ref{LM001}), (\ref{LM002}) and (\ref{LM003}),
\[
\begin{array}{ll}
r_d = r & \Leftrightarrow \#\{X: YH_dX^T=0 \ \ \mbox{for all}\ \ Y \in \mathbb{F}_{p}^{m}\} =p^{m-r} \\
   & \Leftrightarrow \#\{x:\mbox{Tr}(y\phi_d(x))= 0\ \ \mbox{for all}\ \ y\in \mathbb{F}_{q}\} =p^{m-r} \\
      & \Leftrightarrow \phi_d(x) = 0 \ \ \mbox{or}\ \ \phi_d'(x) = 0, \ \ \mbox{has} \ \ p^{m-r} \ \ \mbox{solutions in} \ \ \mathbb{F}_q.
   \end{array}
\]
Thus $r_d\geq 2d$. \qed
\end{proof}

\begin{remark}
 In the above, if the only nonzero element is $\gamma_{t+1}$, and $\gamma_{t+1}^{p^{t+1}}+\gamma_{t+1}=0$, then the corresponding matrix $H_{t+1}$ is a zero matrix. In this case, its rank is zero which does not satisfy that $r_{t+1}\geq 2d$. Hence we assume that $H_d\not= 0_{m\times m}$.
\end{remark}

\begin{lemma} \label{R001}
Let $m\geq 3$ be a positive integer. For $0\leq i\leq \lfloor{{m
}\over 2}\rfloor$ and $\alpha \in \mathbb{F}_q^*$, the rank of the symmetric matrix $H_{\alpha}$ corresponding to the quadratic form $F_{\alpha}(X)=\operatorname{Tr}(f_{\alpha}(x)) = \operatorname{Tr}({\alpha} x^{p^i+1})$
satisfies $r_{\alpha} = m$ or $r_{\alpha}= m-\operatorname{gcd}(2i,m)$. In particular, the rank $r_{\alpha} = m-2i'$ is an even number when $m$ is even and $i'$ is an integer.
\end{lemma}
\begin{proof}
Refer to the proof of Lemma $2$ in \cite{FL001}. \qed
\end{proof}

\subsubsection{Certain results about exponential sums}  \label{Sec3.2.3}

Let $\nu_2(b)$ denote the $2$-adic order function of integer $b$ (i.e., the maximal power of $2$ dividing $b$).
The following two lemmas are about exponential sums (\ref{ES01}) for functions of the form (\ref{QF03}). Lemma \ref{DH001} is for the particular case where only one of $\alpha,\ldots,\beta$ is nonzero. For nonzero $\alpha$, $f_{\alpha}(x)={\alpha} x^{p^j+1}$. Corresponding results are considered in Remark
\ref{R0022} and Remark \ref{R00222} for the cases with $\theta_1^*$ and $\theta_0$ respectively.

 \begin{lemma}(Corollary 7.6., \cite{DH}) \label{DH001}
 Let $\alpha \in \mathbb{F}_{p^m}^*$ and $j\geq 0$.
 \begin{enumerate}
 \renewcommand{\labelenumi}{$($\mbox{\roman{enumi}}$)$}
 \item  If $\nu_2(m)\leq \nu_2(j)$,
 \[
 S(\alpha)=\eta(\alpha)(-1)^{m-1}i^{{1\over 4}(p-1)^2m}p^{{1\over 2}m},
\]
 where $\eta$ is the quadratic character of the finite field $\mathbb{F}_{p^m}$ and $i$ is the imaginary unit.
  \item  If $\nu_2(m)=\nu_2(j)+1$,
 \[
 S(\alpha)=
 \begin{cases}
 p^{{1\over 2}[m+\operatorname{gcd}(2j,m)]}         &\mbox{if}  \ \ {\alpha}^{{{(p^j-1)(p^m-1)}\over {p^{\operatorname{gcd}(2j,m)}-1}}}=-1,\\
 -p^{{1\over 2}m}                          &\operatorname{otherwise}.
 \end{cases}
\]
 \item If $\nu_2(m)>\nu_2(j)+1$,
\[
 S(\alpha)=
 \begin{cases}
 -p^{{1\over 2}[m+\operatorname{gcd}(2j,m)]}         &\mbox{if}  \ \ {\alpha}^{{{(p^j-1)(p^m-1)}\over {p^{\operatorname{gcd}(2j,m)}-1}}}=1,\\
 p^{{1\over 2}m}                          &\mbox{otherwise}.
 \end{cases}
 \]
 \end{enumerate}

  \end{lemma}

Now, exponential sums $R(\alpha,\ldots,\beta)$ can be calculated as in \cite{FL001} that will be needed in the sequel.
\begin{lemma}\label{ES02}
For the quadratic form $F_{\alpha,\ldots,\beta}(X) = XH_{\alpha,\ldots,\beta}X^T$ corresponding to $f_{\alpha,\ldots,\beta}(x)$, see (\ref{QF01}),
\begin{enumerate}
\renewcommand{\labelenumi}{$($\mbox{\roman{enumi}}$)$}
\item
if the rank $r_{\alpha,\ldots,\beta}$ of the symmetric matrix $H_{\alpha,\ldots,\beta}$ is even, which means that $S(\alpha,\ldots,\beta) = \varepsilon p^{m-{r_{\alpha,\ldots,\beta}\over 2}}$, then
\begin{equation}\label{RC01}
R(\alpha,\ldots,\beta) =\varepsilon (p-1) p^{m-{r_{\alpha,\ldots,\beta}\over 2}}; 
\end{equation}
\item
if the rank $r_{\alpha,\ldots,\beta}$ of the symmetric matrix $H_{\alpha,\ldots,\beta}$ is odd, which means that $S(\alpha,\ldots,\beta) = \varepsilon \sqrt{p^*}p^{m-{{r_{\alpha,\ldots,\beta}+1}\over 2}}$, then
\begin{equation}
R(\alpha,\ldots,\beta) = 0
\end{equation}
\end{enumerate}
where $ \varepsilon = \pm 1$.
\end{lemma}

\begin{remark}\label{R0022}
For the exponential sum $S(\alpha,\ldots,\beta)$ corresponding to $f_{\alpha,\ldots,\beta}(x)=\alpha x^{p^i+1}+\cdots+\beta x^{p^j+1}$ with quadratic form $F_{\alpha,\ldots,\beta}(X)$ and symmetric matrix $H_{\alpha,\ldots,\beta}$ (equation (\ref{QF01})), consider $S'(\alpha,\ldots,\beta,\gamma)$ with respect to
\begin{equation}\label{0005}
f'_{\alpha,\ldots,\beta,\gamma}(x)=f_{\alpha,\ldots,\beta}(x)+\gamma x,
\end{equation}
 and $R'(\alpha,\ldots,\beta,\gamma)=\sum_{a=1}^{p-1}S'(a\alpha,\ldots,a\beta,a\gamma)$ (equation (\ref{C01})). From Lemma \ref{FL002}, there are four cases to be considered where the first two equations are for the case with symmetric matrix $H_{\alpha,\ldots,\beta}$ of even rank and the last two equations for the case of odd rank. It can be calculated that
\begin{itemize}
\renewcommand{\labelitemi}{\labelitemiii}
\item
if $S'(\alpha,\ldots,\beta,\gamma)=\varepsilon p^{r'}$, then $R'(\alpha,\ldots,\beta,\gamma)=
\varepsilon (p-1)p^{r'}; $
\item
if $S'(\alpha,\ldots,\beta,\gamma)=\varepsilon \zeta_p^c p^{r'}$, then
$R'(\alpha,\ldots,\beta,\gamma)=
-\varepsilon p^{r'}; $
\item
if $S'(\alpha,\ldots,\beta,\gamma)=\varepsilon \sqrt{p^*}p^{r'}$, then $
R'(\alpha,\ldots,\beta,\gamma)=
0; $
\item
if $S'(\alpha,\ldots,\beta,\gamma)=\varepsilon \zeta_p^c \sqrt{p^*} p^{r'}$, then $
R'(\alpha,\ldots,\beta,\gamma)=
\varepsilon \left({-c}\over p\right)p^{{r'}+{1}}$.
\end{itemize}
In the above, $r'$ is a positive integer, $c \in \mathbb{F}_p^* $, $p^*=\left({-1}\over p\right)p$ and $ \varepsilon =\pm 1$.
\end{remark}

\begin{remark}\label{R00222}
Lemma \ref{ES02} considers the case without $\theta_0, \theta_1^*$, and Remark \ref{R0022} considers the case without $\theta_0$ but with $\theta_1^*$.
\begin{itemize}
\renewcommand{\labelitemi}{\labelitemiii}
\item
 When the primitive idempotent $\theta_0$ is included in the construction of the idempotent of cyclic code, it can be checked from equation (\ref{C02}) that the forms of the results corresponding to Remark \ref{R0022} will not change.
\item
Furthermore, in the weight equation (\ref{C02}), the item $\delta_{b,0}-1$ comes from the consideration of $\theta_0$ comparing with equation (\ref{C01}).
\item
In addition, Lemma \ref{ES02} considers the case of (\ref{0005}) when $\gamma=0$, the results of which will be similar to Remark \ref{R0022} when $\theta_0$ is also included.
\end{itemize}
\end{remark}

\subsection{Cyclotomic cosets and irreducible cyclic codes}  \label{Sec3.3}

In this subsection, properties of cyclotomic cosets and MacWilliams' identities are presented, and the weight distributions of some irreducible cyclic codes corresponding to primitive idempotents in (\ref{PI001}) are investigated. They can be proved by calculating the value distribution of exponential sums or using MacWilliams' identities. These studies will aid the investigations of ODPC in subsequent sections.

 \hspace{0.5cm}

\noindent {\sl Cyclotomic cosets:}

The cyclotomic coset containing $s$ is defined to be $\mathcal{D}_s=\{s,sp,sp^2,\ldots,sp^{m_s-1}\}$
where $m_s$ is the smallest positive integer such that $p^{m_s}\cdot s \equiv s \ (\mbox{mod} \ p^m-1)$.
The primitive idempotent $\theta_s$ is defined as a polynomial satisfying $\theta_s(\pi^j)=1, \ \mbox{if} \ \ j\in \mathcal{D}_s, \ \mbox{and} \ 0 \ \mbox{otherwise}$
where $\pi$ is a primitive element of the field $\mathbb{F}_q=\mathbb{F}_{p^m}$. Also, define the primitive idempotent $\theta_s^*$ by
$\theta_s^*(\pi^j)=1, \ \mbox{if} \ \ j\in \mathcal{D}_{-s}, \ \mbox{and} \ 0 \ \mbox{otherwise}$.

In this paper, we will need primitive idempotents of the forms $\theta_0, \theta_1^*, \theta_{l_i}^*$, where $l_i=1+p^i, \ \ i=0, 1, 2, \ldots, \lfloor {m\over 2}\rfloor$. {\bf The irreducible cyclic code with primitive idempotent $\theta_{l_i}^*$ is denoted by $\mathcal{C}_i$. In addition, $\mathcal{C}_0'$ corresponds to the case of $\theta_1^*$.}
 The following lemma is about the size of cyclotomic cosets. 
\begin{lemma}\label{GCD002}
If $m=2t+1$ is odd, then for $l_i=1+p^i$, the cyclotomic coset $\mathcal{D}_{l_i}$ has size
\[
|\mathcal{D}_{l_i}|=m, \ \ \ 0\leq i \leq t.
\]
If $m=2t+2$ is even, then for $l_i=1+p^i$, the cyclotomic coset $\mathcal{D}_{l_i}$ has size
\[
|\mathcal{D}_{l_i}| =
\begin{cases}
m, & 0\leq i \leq t \\
m/2, & i=t+1.
\end{cases}
\]
\end{lemma}

 \hspace{0.5cm}

\noindent {\sl MacWilliams' identities:}

 In \cite {MS001}, there is MacWilliams' theorem for Hamming weight enumerators of linear codes over finite field.
Let $A_i$ be the number of codewords of weight $i$ in a code $C$ with length $l$ and dimension $k$ where $0\leq i\leq l$. Let $A_i'$ be the corresponding number in the dual code $C^{\perp}$ with dimension $l-k$. Then
 \begin{equation}\label{MS01}
W_C(x,y)={1\over |C^{\perp}|}W_{C^{\perp}}(x+(p-1)y,x-y)
\end{equation}
where $W_C(x,y) = \sum_{i=0}^{l}{A_ix^{l-i}y^i}$. Setting $x=1$, from (\ref{MS01}) we deduce
\begin{equation}\label{MS002}
\sum_{i=0}^{l}A_iy^i = {1\over {p^{l-k}}}\sum_{i=0}^{l}A_i'(1+(p-1)y)^{l-i}(1-y)^i.
\end{equation}
After differentiating (\ref{MS002}) with respect to $y$ and setting $y=1$, the first Pless moment identity is obtained for $l\geq 2$
\begin{equation}\label{MS003}
\sum_{i=1}^{l}{{iA_i}\over{p^k}}={1\over p}((p-1)l-A_1')={1\over p}(p-1)l \ \ \ \mbox{if} \ \ A_1'=0.
\end{equation}
Differentiating again, we can get the second Pless moment identity for $l\geq 3$, etc.

 \hspace{0.5cm}

\noindent {\sl Weights of some irreducible cyclic codes:}
\begin{lemma}\label{SC01}
Let $m>1$ be an integer. The irreducible cyclic code $\mathcal{C}_0'$ with primitive idempotent $\theta_1^*$ has only one nonzero weight $p^{m-1}(p-1)$.
\end{lemma}

The following lemma is obtained from equation (\ref{C01}) by using Lemma \ref{DH001} and Lemma \ref{ES02} where $R(\alpha)=\pm (p-1)p^{{m\over 2}+u}$.
\begin{lemma} \label{C001}
 Let $m=2t+2, t\geq 1$. For $0\leq i\leq t$, the irreducible cyclic code $\mathcal{C}_i$ with primitive idempotent $\theta_{l_i}^*$ has weights of the forms $p^{m-1}(p-1)\pm (p-1)p^{{m\over 2}+u-1}$ where $u=0$ or $i'$ defined in Lemma \ref{R001}.
\end{lemma}

In fact, $p^{m-1}(p-1)+(p-1)p^{{m\over 2}+u-1}$ corresponds to the negative value of $S(\alpha)$ in Lemma \ref{DH001}, and $p^{m-1}(p-1)-(p-1)p^{{m\over 2}+u-1}$ corresponds to the positive value of $S(\alpha)$.

\begin{lemma}\label{T001}
Let $m=2t+2, t\geq 1$. The irreducible cyclic code $\mathcal{C}_{t+1}$ with primitive idempotent $\theta_{l_{t+1}}^*$ has only one nonzero weight, and $A_{p^{m-1}(p-1)+{{p-1}\over p}p^{m\over 2}} = p^{m\over 2}-1$.
\end{lemma}

\begin{example}
Set $m=8,p=3$. The irreducible cyclic code $\mathcal{C}_4$ with primitive idempotent $\theta_{l_4}^*$ has generator polynomial
\[
g(x)={{x^{6560}-1}\over {x^4+x^3-1}}
\]
and only one nonzero weight $4428$.
\end{example}

\begin{lemma}\label{T002}
Let $m=2t+2, t\geq 1$. The irreducible cyclic code $\mathcal{C}_{0}$ with primitive idempotent $\theta_{l_{0}}^*$ has two nonzero weights, and $A_{p^{m-1}(p-1)+{{p-1}\over p}p^{m\over 2}} = A_{p^{m-1}(p-1)-{{p-1}\over p}p^{m\over 2}}={{p^{m }-1}\over 2}$.
\end{lemma}

  Lemma \ref{L002} is about the weight distributions of a class of irreducible cyclic codes with $m$ in the form of double factorial. As for Lemma \ref{C001}, it can be deduced using Lemma \ref{DH001}.
\begin{lemma}\label{L002}
Let $m=2t+2=1\cdot 3\cdot 5\cdot 7\cdots (2t'-1)\cdot {2^{s_0}}$ where $t'\geq 3$ and  $s_0\geq 2$. For $i=1\cdot 2^{s_0-1},3\cdot 2^{s_0-1},5\cdot 2^{s_0-1},7\cdot 2^{s_0-1},\ldots,(2t'-1)\cdot 2^{s_0-1}$, the irreducible cyclic code $\mathcal{C}_i$ with primitive idempotent $\theta_{l_i}^*$ has two nonzero weights and
\[
A_{p^{m-1}(p-1)+{{p-1}\over p}p^{m\over 2}} = {p^i\over {p^i+1}}\left(p^m-1\right), \ \ A_{p^{m-1}(p-1)-{{p-1}\over p}p^{{m\over 2}+i}} = {1\over {p^i+1}}\left(p^m-1\right).
\]
\end{lemma}

\subsection{The ODPCs-II$^{inv}$ of a class of cyclic codes} \label{Sec3.5}

Before going on, let's consider the following lemma
which is the counterpart of Lemma \ref{MS02} for evaluating the ranks of corresponding quadratic forms in the third step of the proof of Theorem \ref{F01}.

For $f_d'(x) =\alpha_0 x^2+\alpha_1x^{p+1}+\cdots+\alpha_dx^{p^d+1}$ with corresponding quadratic form $F_d'(X)=\mbox{Tr}(f_d'(x))=XH_d'X^T$ where $(\alpha_0,\alpha_1,\ldots,\alpha_d)\in \mathbb{F}_q^{d+1}\backslash\{(0,0,\ldots,0)\}$, the following result is about its rank.
\begin{lemma}\label{RQ002}
Let $m\geq 3$ be a positive integer, $0\leq d\leq \lfloor{m\over 2}\rfloor$. The rank $r_d'$ of the symmetric matrix $H_d'$ satisfies $r_d'\geq m-2d$.
\end{lemma}

\begin{proof}
Refer to the poof of Lemma \ref{MS02} where the counterpart here means that the lowest degree in Lemma \ref{MS02} is the highest degree in this lemma. \qed
\end{proof}

\begin{theorem}\label{F01}
Let $m=2t+2=1\cdot 3\cdot 5\cdot 7\cdots (2t'-1)\cdot {2^{s_0}}$ where $t'\geq 3$ and  $s_0\geq 2$. For
the cyclic code ${C}$ with idempotent
\[
\theta_0+\theta_1^*+\theta_{l_0}^*+\theta_{l_{1\cdot 2^{s_0-1}}}^*+\theta_{l_{3\cdot 2^{s_0-1}}}^*+\theta_{l_{5\cdot 2^{s_0-1}}}^*+\cdots+ \theta_{l_{(2t'-1)\cdot 2^{s_0-1}}}^*,
\]
the following order of adding the primitive idempotents one-by-one
\begin{equation}\label{C002}
\theta_0,\theta_1^*,\theta_{l_0}^*,\theta_{l_{1\cdot 2^{s_0-1}}}^*,\theta_{l_{3\cdot 2^{s_0-1}}}^*,\theta_{l_{5\cdot 2^{s_0-1}}}^*,\cdots, \theta_{l_{(2t'-1)\cdot 2^{s_0-1}}}^*
\end{equation}
 will get the ODPC-II$^{inv}$.
The corresponding minimum distances of the above chain satisfy that
\[
\begin{array}{l}
d_{\tau_{t'+2}}=p^m-1,\\
d_{\tau_{t'+1}}=p^{m-1}(p-1)-1 \ \ \mbox{and} \\
 d_{\tau_{t'-d+3}}=p^{m-1}(p-1)-1-{{p-1}\over p}p^{{m\over 2}+{(2(d-3)-1)\cdot 2^{s_0-1}\cdot (1-\delta_{d,3})}} \ \ \mbox{for} \ \ 3\leq d\leq t'+3
\end{array}
\]
 where $\delta_{d,3}=1$ for $d=3$ and $0$ otherwise.
\end{theorem}

\begin{proof}
To get the ODPC-II$^{inv}$, the selection of subcodes from the first to the end can be proceeded by adding the primitive idempotents one-by-one in the following order.
\begin{enumerate}
\renewcommand{\labelenumi}{$($\mbox{\roman{enumi}}$)$}
\item
In the first step, primitive idempotent $\theta_0$ corresponds to the cyclic subcode ${C}_{\tau_{t'+2}}$ with all one vector and its multiples. It is obvious that it has the largest minimum distance $d_{\tau_{t'+2}}=p^m-1$.
\item
In the second step, if the primitive idempotent $\theta_1^*$ is selected, the corresponding cyclic subcode ${C}_{\tau_{t'+1}}$ with idempotent
\begin{equation}\label{0001}
\theta_0+\theta_1^*
\end{equation}
 has minimum distance $d_{\tau_{t'+1}}=p^{m-1}(p-1)-1$ according to Lemma \ref{SC01} and equation (\ref{C02})
 \[
w_H(c) =p^{m-1}(p-1)+\delta_{b,0}-1-{1\over p}\sum_{a=1}^{p-1}{\zeta_p^{ab}}S(a\alpha_1,\ldots,a\alpha_{\iota})
\]
where $\delta_{b,0}$ is $1$ for $b=0$, and $0$ otherwise.
 It is bigger than the minimum distance of any cyclic code obtained by replacing $\theta_1^*$ with $\theta_{l_{i}}^*$ in (\ref{0001}). In fact, any irreducible cyclic subcode $\mathcal{C}_{ {i}}$ with primitive idempotent $\theta_{l_{i}}^*$ has distance $p^{m-1}(p-1) -(p-1)p^{{m\over 2}+i-1}$
  by Lemma \ref{T002} and Lemma \ref{L002}, where $i=0,1\cdot 2^{s_0-1},3\cdot 2^{s_0-1},\ldots,(2t'-1)\cdot 2^{s_0-1}$.

\item
For the $d$th step, where $3\leq d\leq t'+3$,
the cyclic code $C_{\tau_{t'-d+3}}$ obtained from (\ref{C002}) has minimum distance
\[
d_{\tau_{t'-d+3}}=p^{m-1}(p-1)-1-{{p-1}\over p}p^{{m\over 2}+{(2(d-3)-1)\cdot 2^{s_0-1}\cdot (1-\delta_{d,3})}}
\]
 where $\delta_{d,3}=1$
 for $d=3$ and $0$ otherwise, see follows.
\begin{itemize}
\renewcommand{\labelitemi}{\labelitemiii}
\item
 In fact, Lemma \ref{T002} and Lemma \ref{L002} imply that there is a codeword $c$ of weight
 \begin{equation}\label{0002}
 d_{\tau_{t'-d+3}}+1
 \end{equation}
  in the irreducible cyclic code $\mathcal{C}_{u}$ with primitive idempotent $\theta_{l_{u}}^*$, which is a subcode of ${C}_{\tau_{t'-d+3}}$ where $u={(2(d-3)-1)\cdot 2^{s_0-1}\cdot (1-\delta_{d,3})}$.
\item
 In addition, by Lemma \ref{RQ002}, the ranks of quadratic forms of the codewords in the cyclic subcode ${C}_{\tau_{t'-d+3}}$ are not less than
 $ m-2u.$
 \item
  According to Lemma \ref{ES02}, the rank of the quadratic form of the codeword with weight (\ref{0002}) is $m-2u$ which doesn't depend on $\theta_0, \theta_1^*$. And then we can find another codeword with actual minimum weight $d_{\tau_{t'-d+3}}$ of ${C}_{\tau_{t'-d+3}}$ by adding a linear polynomial to $c$, see the second item of Remark \ref{R00222} in the case of $\delta_{b,0}=0$.
\end{itemize}

 Any other selection of cyclic subcode with $d$ primitive idempotents (\ref{C002})
will contain a codeword of weight strictly less than $d_{\tau_{t'-d+3}}$ by Lemma \ref{T002} and Lemma \ref{L002}. \qed

\end{enumerate}

\end{proof}

\begin{remark}\label{MF0001}
Suppose the parameters $m,t,t'$ and $s_0$ satisfy the conditions of Theorem \ref{F01}. For a cyclic code corresponding to the sum of some primitive idempotents
\begin{equation}\label{R004}
\theta_0,\theta_1^*,\theta_{l_{0}}^* \ \ \mbox{and} \ \ \theta_{l_{i\cdot 2^{s_0-1}}}^*,\ldots, \theta_{l_{j\cdot 2^{s_0-1}}}^*
\end{equation}
where $i,\ldots,j$ are divisors of $1\cdot 3\cdot 5\cdot 7\cdots (2t'-1)$, the ODPC-II$^{inv}$ can be analyzed in a similar process.

Furthermore, the ODPC can be studied for many natural numbers $m$ with $\nu_2(m)\geq 2$ according to the factorization of $m$. In fact, for $
m=p_1^{n_1}p_2^{n_2}\cdots p_{t''}^{n_{t''}}2^{s_0}$
 where $p_1,p_2,\ldots,p_{t''}$ are odd primes,
 the parameters $i,\ldots,j$ of formula (\ref{R004}) lie in the set
\[
\{ p_1^{n_1'}p_2^{n_2'}\cdots p_{t''}^{n_{t''}'}|0\leq n_1'\leq n_1,\ldots,0\leq {n_{t''}'} \leq {n_{t''}} \}
\]
which has size
$N=(n_1+1)(n_2+1)\cdots (n_{t''}+1).$
  Considering the first three elements of (\ref{R004}), the total number of cyclic codes, the ODPC-II$^{inv}$ of which can be obtained by using a similar method of Theorem \ref{F01}, is about $2^{N+2}$.
\end{remark}

\section{Bounds and achievability on the ODPC of $\mathcal{GRM}(2,m)^*$} \label{Sec4}

This section investigates the ODPC of $\mathcal{GRM}(2,m)^*$ in the inverse dictionary order. In fact, we are trying to extend a cyclic subcode chain by increasing the length of information bits and keep the minimum distances as large as possible in each step. For this, necessary knowledge about alternating bilinear forms and quadratic forms is discussed in Section \ref{Sec4.1}. The possible values of the parameter $c$ in Lemma \ref{FL002} are investigated in Section \ref{Sec4.3}, which analyzes the forms of the minimum distances. Then in Section \ref{Sec4.4}, the bounds and achievability on ODPC are analyzed separately according to $m$ is even or odd under Standard I and Standard II, see  Theorem \ref{ME002}, Theorem \ref{EP02}, Proposition \ref{E2002} and  Proposition \ref{E2001}.

\subsection{Alternating bilinear forms and quadratic forms}\label{Sec4.1}

In this subsection, a brief explanation and properties of alternating bilinear forms, quadratic forms, skew-symmetric matrices, symmetric matrices, and $(m,d)$-sets are presented.\\

\noindent {\sl Alternating bilinear forms}:

Let $V$ be an $m$-dimensional vector space over the field $\mathbb{F}_p$ where $p$ is an odd prime. An alternating bilinear form on $V$ is a bilinear form $B(\cdot,\cdot)$ which satisfies
\[
B(\mathbf{x},\mathbf{x})=0,
\]
from which it follows that
\[
B(\mathbf{x},\mathbf{y}) + B(\mathbf{y},\mathbf{x}) = 0\ \ \forall \mathbf{x,y} \in V.
\]

\begin{remark}
When $p=2$, it is also called symplectic form as defined in \cite{MS001}. A more general definition over a finite field $\mathbb{F}_q$ can also be stated, where $q$ is a power of a prime.
\end{remark}

A skew-symmetric matrix $B = [b_{i,j}]$ of order $m$ is a matrix which satisfies
\[
b_{i,i}=0, \ \ \ b_{i,j}+b_{j,i} = 0.
\]
 There is a one-to-one correspondence between the set $\mathbb{B}(m,p)$ of alternating bilinear forms on $V$ and the set (denoted by $Y_m=Y(m,p)$) of skew-symmetric matrices of order $m$ over $\mathbb{F}_p$. Clearly, $Y_m$ is an
${m(m-1)\over 2}$-dimensional vector space over $\mathbb{F}_p$.
The rank of a form $B(\cdot,\cdot)$ is the rank of its skew-symmetric matrix, which is even.

Set $n=\lfloor m/2\rfloor$.
For $k=0,1,\ldots,n$, the partition $R' =\{R_0',R_1',\ldots,R_n' \}$ of $Y_m^2=Y_m \times Y_m$ is defined by
\[
R_k' =\{(A,B)\in Y_m^2|\mbox{rank}(A-B)=2k\}.
\]
Clearly, $R_k'$ is symmetric binary relation on $Y_m$, and $R_0'$ is the diagonal relation. In fact $(Y_m,R')$ is an association scheme with $n$ classes, see \cite{DG01}. The reader is referred to \cite{CV,D02,MS001,S01} for
a definition and fundamental properties of association schemes.

An $(m,d)$-set $Y$ is a subset of $Y_m$ satisfying that
\begin{equation}\label{SS01}
\mbox{rank}(A-B)\geq 2d, \ \ \ \forall A,B \in Y, \ \ \ A\not= B,
\end{equation}
where $1\leq d\leq n$. Set $c = p^{m(m-1)/{2n}}$. The following result is about the Singleton bound on the size of $(m,d)$-set.
\begin{lemma}(Theorem 4., \cite{DG01})\label{MS04}
For any $(m,d)$-set $Y$, we have (the Singleton bound)
$|Y| \leq c^{n-d+1}$.
 \end{lemma}


\noindent {\sl Quadratic forms}:

Let $\mathbb{Q}(m,p)$ be the set of all quadratic forms of an $m$-dimensional vector space $V$ over $\mathbb{F}_p$, where $p$ is an odd prime. Then $\mathbb{Q}(m,p)$ can be considered as a vector space of dimension $m(m+1)/2$ over $\mathbb{F}_p$. In fact, there is a one-to-one correspondence from $\mathbb{Q}(m,p)$ to the set (denoted by $X_m=X(m,p)$) of symmetric matrices with order $m$ over $\mathbb{F}_p$.
The more general definition of quadratic forms is over an arbitrary finite field $\mathbb{F}_q$ where $q$ is a power of a prime.

Let
\[
R = \{R_i:i=0,1,2,\ldots, \lfloor {{m+1}\over 2}\rfloor\}
\]
be the set of symmetric relations $R_i$ on $X_m$ defined by
\[
R_i = \{ (A,B)|A,B \in X_m, \mbox{rank}(A-B)=2i-1 \ \mbox{or} \ 2i   \}.
\]
It is easy to verify that $R$ is a partition of $X_m^2$.

Corresponding to the association scheme $(Y_m,R')$ of skew-symmetric matrices, the following results are about the scheme $(X_m,R)$ in \cite{E01}.

\begin{lemma} (Theorem 1., \cite{E01})
$(X_m,R)$ forms an association scheme of class $\lfloor(m+1)/2\rfloor$.
\end{lemma}

\begin{lemma}(Theorem 2., \cite{E01})\label{PQS}
All the parameters (and consequently all the eigenvalues) of the two association schemes $(X_m,R)$ and $(Y_{m+1},R')$ of class $\lfloor(m+1)/2\rfloor$ are exactly the same.
\end{lemma}

A set $X\subset X_m$ is called an $(m,d)$-set of $X_m$ if it satisfies
\[
\mbox{rank}(A-B)\geq 2d-1, \ \ \ \forall A\not= B \in X_m,
\]
where $1\leq d \leq \lfloor(m+1)/2\rfloor$. Note that, the $(m,d)$-set of $Y_m$ implies that $\mbox{rank}(A-B)\geq 2d$ in (\ref{SS01}).
The following proposition is about the size of the $(m,d)$-set of $X_m$, which is a counterpart of Lemma \ref{MS04}.
\begin{proposition}\label{EF002}
 Set $c=p^{m(m+1)/{2n}}, n = \lfloor(m+1)/2\rfloor$.
For any $(m,d)$-set $X\subset X_m$, we have (the Singleton bound) $|X| \leq c^{n-d+1}$.
\end{proposition}

\subsection{The parameter $c$ in Lemma \ref{FL002}} \label{Sec4.3}

In this subsection, the possible values of the parameter $c$ in Lemma \ref{FL002} are investigated in Lemma \ref{LE04} for rank $r\geq 2$.
In fact, using Corollary \ref{CEP001} and Lemma \ref{E06}, any element $c$ of $\mathbb{F}_p$ can be constructed by (\ref{EC01}) in Lemma \ref{FL002}. This is necessary for us to understand the weights of some corresponding codewords, especially the minimum weights. Lemma \ref{R03} considers the codeword weights for the case of rank $1$.

\begin{lemma}\label{EP001}
Every element in a finite field $\mathbb{F}_q$ can be written as a sum of two squares where $q=p^m$ for some prime $p$.
\end{lemma}

\begin{corollary}\label{CEP001}
Let $f(x,y,\ldots,z)=\alpha x^2+\beta y^2+\cdots +\gamma z^2$ where $\alpha, \beta, \ldots, \gamma \in \mathbb{F}_q$ are fixed and at least two of the coefficients are nonzero.
Then every element $a$ of $\mathbb{F}_q$ can be expressed by $f(x,y,\ldots,z)$, i.e., there are $x_a,y_a,\ldots,z_a\in \mathbb{F}_q$ such that $a=f(x_a,y_a,\ldots,z_a)$.
\end{corollary}

\begin{proof}
It is only necessary to consider the situation such that, for any given  $f(x,y)=\alpha x^2+\beta y^2 (\alpha,\beta \not=0)$, there are $x_a,y_a\in \mathbb{F}_q$ satisfying $a=f(x_a,y_a)$, see the following three subcases.
\begin{enumerate}
\renewcommand{\labelenumi}{$($\mbox{\roman{enumi}}$)$}
\item  If $\alpha$ and $\beta$ are both squares, then $\alpha ={\alpha_0}^2$ and $\beta={\beta_0}^2$. Therefore
\[
f(x,y)=(\alpha_0x)^2+(\beta_0y)^2=x'^2+y'^2 \mbox{where} \ \ x'=\alpha_0x, y'=\beta_0y.
\]
According to Lemma \ref{EP001}, every element of $\mathbb{F}_q$ can be expressed by $f(x,y)$.
\item  If $\alpha$ and $\beta$ are both non-squares, then $\alpha =c_0{\alpha_0}^2$ and $\beta=c_0{\beta_0}^2$ where $c_0\in \mathbb{F}_q^*$ is a non-square element. Thus
\[
\begin{array}{ll}
f(x,y)&=c_0\left((\alpha_0x)^2+(\beta_0y)^2\right)=c_0f'(x,y)
\end{array}
\]
where $f'(x,y)=x'^2+y'^2$, $x'=\alpha_0x$ and $y'=\beta_0y$.
From Lemma \ref{EP001}, every element of $\mathbb{F}_q$ can be expressed by $f'(x,y)$, and then by $f(x,y)$.
\item If $\alpha$ is a square and $\beta$ is a non-square, then $\alpha ={\alpha_0}^2$ and $\beta=c_0{\beta_0}^2$, where $c_0\in \mathbb{F}_q^*$ is a non-square element. Take $y=0$, every square element of $\mathbb{F}_q$ can be expressed by $f(x,0)=(\alpha_0x)^2$. Take $x=0$, every non-square element of $\mathbb{F}_q$ can be expressed by $f(0,y)=c_0(\beta_0y)^2$.  \qed
\end{enumerate}
\end{proof}

 For $f_{\alpha,\ldots,\beta,\gamma}'(x)=\alpha x^{p^i+1}+\cdots+\beta x^{p^j+1}+\gamma x$ with domain $\mathbb{F}_q$, consider the corresponding exponential sum of the form $S'(\alpha,\ldots,\beta,\gamma)=\sum_{X\in \mathbb{F}_p^m}{\zeta_p}^{XHX^T+AX^T}$ as defined in Remark \ref{R0022}, where $\alpha,\ldots,\beta,\gamma \in \mathbb{F}_q, q=p^m$. Lemma \ref{E06} states that the vector $A$ can be all elements of $\mathbb{F}_p^m$, and then combines with Corollary \ref{CEP001} to support Lemma \ref{LE04}.

\begin{lemma}\label{E06}
Let $f_{\gamma}'(x)=\gamma x$, then
\begin{equation}\label{CC001}
\begin{array}{ll}
 \operatorname{Tr}(f_{\gamma}'(x)) =\operatorname{Tr}(\gamma x)
                      =AX^T
\end{array}
\end{equation}
 where $A=(\operatorname{Tr}(\gamma v_1),\cdots,\operatorname{Tr}(\gamma v_m))\in \mathbb{F}_p^m$, $x=x_1v_1+\cdots+x_mv_m$, $x_i\in \mathbb{F}_p$ and $v_1,\ldots,v_m$ is a basis of $\mathbb{F}_q$ over $\mathbb{F}_p$. When $\gamma$ varies over the elements of $\mathbb{F}_q$, $A$ may be any element of $\mathbb{F}_p^m$.
\end{lemma}

\begin{proof}
From (\ref{CW001}) in the case of $\iota =1$, it can be checked that $\mbox{Tr}(f_{\gamma}'(x))$ generates all the codewords of $\mathcal{C}_0'$ with primitive idempotent $\theta_1^*$, whose dimension is $m$. In addition, the number of codewords and the number of $\gamma s$ are the same as $p^m$.
So different $\gamma$ represents different codeword in (\ref{CC001}). In another word, different $A s$ represent $p^m$ different codewords. \qed
\end{proof}

\begin{lemma} \label{LE04}
The parameter $c$ in Lemma \ref{FL002} can be any element of $\mathbb{F}_p$ if the rank $r$ of corresponding quadratic form $F(X)$ is not less than $2$.
\end{lemma}

\begin{proof}
It is known that there exists a matrix $M\in \mbox{GL}{\mathbb{F}_p^m}$ such that $H'=MHM^T$ is a diagonal matrix $H'=\mbox{diag}(a_1,a_2,\ldots,a_r,0,\ldots,0)$ where $a_i\in \mathbb{F}_p^* (1\leq i\leq r)$. In addition, for any $c\in \mathbb{F}_p$, there exists a relation between $c, A, B$ and matrix $H$ as stated in the following from Lemma \ref{FL002},
\begin{equation}
\begin{array}{ll}\label{CEP002}
c&={1\over 2}AB^T\\
 &={1\over 2}(-2BHB^T)=-BHB^T=-B'H'B'^T\\
 &=-(a_1x_1^2+\cdots+a_rx_r^2)
\end{array}
\end{equation}
where $B$ is a variable vector and $B'=BM^{-1}$.
Furthermore, from Corollary \ref{CEP001} there exists $B'=(x_1,\ldots,x_r,0,\ldots,0)$ satisfying equation (\ref{CEP002}),
then $B=B'M$. In this case $A=-2BH$ which from Lemma \ref{E06} corresponds to an element $\gamma \in \mathbb{F}_q$. \qed
\end{proof}

The following lemma is about the possible codeword weights when the symmetric matrix has rank $1$.
\begin{lemma}\label{R03}
For the function $f_{\alpha,\ldots,\beta}(x)$ defined in (\ref{QF03}), denote its corresponding quadratic form and symmetric matrix by $F_{\alpha,\ldots,\beta}(X)$ and $H_{\alpha,\ldots,\beta}$ respectively. Suppose the rank of the matrix satisfies $r_{\alpha,\ldots,\beta}=1$. Then for the function $f_{\alpha,\ldots,\beta,\gamma}(x)=f_{\alpha,\ldots,\beta}(x)+\gamma x$, the weights of corresponding codewords (see, (\ref{C01})) are $p^{m-1}(p-1)$ and $p^{m-1}(p-2)$ when $\gamma$ varies over $\mathbb{F}_q$.
\end{lemma}

\begin{proof}
According to Lemma \ref{FL002}, the exponential sum $S'({\alpha,\ldots,\beta,\gamma})$ corresponding to $f_{\alpha,\ldots,\beta,\gamma}(x)$ has value $\zeta_{p}^{c}\left(h\over p\right)\sqrt{p^*}p^{m-1}$, where $c=-b^2h$, $b\in \mathbb{F}_p$ and $h$ is the nonzero element of the one-dimensional matrix $H'$ as stated in the paragraph before Lemma \ref{FL002}. By Remark \ref{R0022}, the corresponding exponential sum $R'({\alpha,\ldots,\beta,\gamma})$ is $0$ if $b=0$, and $\left(h\over p\right) \left({b^2h}\over p\right)p^{m}=p^m$ if $b\not=0$. From the relation between the weight of codeword and corresponding exponential sum (\ref{C01}), this lemma is obtained. \qed
\end{proof}

\subsection{Main results on ODPC of $\mathcal{GRM}(2,m)^*$} \label{Sec4.4}

 In this subsection, bounds and achievability on the ODPC are studied for the case of $m=2t+2$ in Section \ref{Sec4.4.1} and for the case of $m=2t+1$ in Section \ref{Sec4.4.2}. Theorem \ref{ME002} is for Standard I and even $m$, and Theorem \ref{EP02} is for Standard II and even $m$. The case of odd $m$ under Standard I appears in Proposition \ref{E2002}, and the case of odd $m$ under Standard II is provided in Proposition \ref{E2001}.

\subsubsection{The case of $m=2t+2$}\label{Sec4.4.1}

In this subsection the ODPC of the cyclic code $\mathcal{GRM}(2,m)^*$ is studied in the inverse dictionary order, which has idempotent $\theta_0+\theta_1^*+\theta_{l_{t+1}}^*+\cdots+\theta_{l_1}^*+\theta_{l_0}^*$, see Section \ref{Sec3.1}. According to Lemma \ref{th1}, the length of the cyclic subcode chains is $\lambda=t+4$, the number of chains is $(t+4)!$, the number of chains contained in each class is $\mu = (t+2)!1!1!=(t+2)!$, and the number of classes is ${{\lambda !}\over {\mu}}=(t+4)(t+3)$.

As to the comparison between the distance profiles under Standard I, when increasing the message length in the cyclic subcode chain, we focus on the decreasing dimension profile here
\begin{equation}\label{DPF001}
\begin{array}{l}
 (t+2)m+{m\over 2}+1, \ldots, jm+{m\over 2}+1, (j-1)m+{m\over 2}+1,\\
(j-1)m+{m\over 2}, \ldots, (i+1)m+{m\over 2}, im+{m\over 2},\\
im, \ldots, 2m, m.
\end{array}
\end{equation}
It means that the primitive idempotent $\theta_{l_{t+1}}^*$ is accumulated in the $(i+1)$th order to construct the cyclic code with dimension $im+{m\over 2}$ in Theorem \ref{ME002}, and $\theta_0$ is accumulated in the $(j+1)$th order to construct the cyclic code with dimension $(j-1)m+{m\over 2}+1$, where $2\leq i<j\leq t+1$.

\begin{remark}
  The focus on equation (\ref{DPF001}) is of universal significance, where integer $j$ is selected satisfying $3\leq j\leq t+1$. For any such $j$, the number of $i$ that can be selected is $j-2$. Altogether the number of classes included by equation (\ref{DPF001}) is
\[
\sum_{j=3}^{t+1}(j-2)={{t(t-1)}\over 2}
\]
which is almost half of the whole number of classes $(t+4)(t+3)$.
\end{remark}

The following $t+4$ sets will restrict the minimum distances in Theorem \ref{ME002}.
\begin{itemize}
\renewcommand{\labelitemi}{\labelitemiii}
\item First part:
\begin{eqnarray}
B_{\tau_{t+3}}&=&\{p^{m-1}(p-1)\}; \nonumber \\
B_{\tau_{t+2}}&=&\{p^{m-1}(p-1)-(p-1)p^{t+1}, p^{m-1}(p-1)-p^{t+1} \nonumber \\
  && p^{m-1}(p-1)-(p-1)p^{t}, p^{m-1}(p-1)-p^{t}\}; \nonumber
\end{eqnarray}
\item Second part: for $3\leq d\leq i$,
\begin{equation}\label{B01}
\begin{array}{ll}
B_{\tau_{t-d+4}}=&\{p^{m-1}(p-1)-p^{t+d-2}, p^{m-1}(p-1)-(p-1)p^{t+d-2},\\
& p^{m-1}(p-1)-p^{t+d-1}, p^{m-1}(p-1)-(p-1)p^{t+d-1}\};
\end{array}
\end{equation}
\item Third part: for $i+1\leq d\leq j$,
\[
  B_{\tau_{t-d+4}}=\{p^{m-1}(p-1)-p^{t+d-2}, p^{m-1}(p-1)-(p-1)p^{t+d-2}\};
\]
\item Fourth part: for $j+1\leq d\leq t+3$,
\[
  B_{\tau_{t-d+4}}=\{p^{m-1}(p-1)-1-(p-1)p^{t+d-3}, p^{m-1}(p-1)-1-p^{t+d-3}\};
\]
\item Final part: $B_{\tau_{0}}=\{p^{m-1}(p-2)-1\}$.
\end{itemize}

\begin{theorem}\label{ME002}
Let $m=2t+2, t\geq 2$. For the cyclic code $\mathcal{GRM}(2,m)^*$,
consider the dimension profile (\ref{DPF001}) under standard I.
 The chain obtained by accumulating the primitive idempotents
 one-by-one in the following order
\begin{equation}\label{CDP002}
\theta_1^*,\theta_{l_t}^*,\ldots,\theta_{l_{t-i+2}}^*,\theta_{l_{t+1}}^*,\theta_{l_{t-i+1}}^*,\ldots,\theta_{l_{t-j+3}}^*,\theta_{0},\theta_{l_{t-j+2}}^*,\ldots,\theta_{l_1}^*,\theta_{l_0}^*
\end{equation}
 provides a distance profile satisfying
\begin{equation}\label{CDP003}
d_{\tau_u}\in B_{\tau_u} (0\leq u\leq t+3)
\end{equation}
 in the inverse dictionary order, where $2\leq i<j\leq t+1$.
 The corresponding cyclic subcode chain (expansion from right to left) is denoted by
\[
 C_{\tau_{0}}\supset C_{\tau_{1}}\supset \cdots \supset C_{\tau_{t+2}}\supset C_{\tau_{t+3}}.
\]
For any other cyclic subcode chain, the minimum distance is less than or equal to the corresponding $\max B_{\tau_u}$, especially the ODPC.
\end{theorem}

\begin{proof}
The following proof is about how to expand the cyclic subcodes by accumulating the primitive idempotents in (\ref{CDP002}) with corresponding minimum distances from the sets $B_{\tau_u} (t+3\geq u\geq 0)$ in the five parts listed before Theorem \ref{ME002}.

By Lemma \ref{GCD002}, the dimension of the irreducible cyclic code $\mathcal{C}_i$ is $m$ for $0\leq i \leq t$, $\dim\mathcal{C}_{t+1}={m\over 2}$, $\dim\mathcal{C}_{-1}=1$ and $\dim\mathcal{C}_0'=m$ where $\mathcal{C}_{-1}$ and $\mathcal{C}_0'$ correspond to $\theta_0$ and $\theta_1^*$ respectively. Since the dimension profile is of the form (\ref{DPF001}) by the assumption under Standard I, primitive idempotents $\theta_{l_{t+1}}^*$ (of $\mathcal{C}_{t+1}$) and $\theta_0$ are selected in the $(i+1)$th and $(j+1)$th orders respectively.

  Let's consider the first $i$ orders in the selection process.

{\bf Firstly}, from Lemma \ref{SC01}, Lemma \ref{C001} and Lemma \ref{T001}, except for $\theta_{l_{t+1}}^*$, only the primitive idempotent $\theta_1^*$ corresponding to the cyclic code ${C}_{\tau_{t+3}}$ can achieve the optimum minimum distance $d_{\tau_{t+3}}=p^{m-1}(p-1)$. Then  $d_{\tau_{t+3}}\in B_{\tau_{t+3}}$.
For the second order using the set $B_{\tau_{t+2}}$, it can be checked that the minimum distance of the cyclic code ${C}_{\tau_{t+2}}$ with idempotent $\theta_1^*+\theta_{l_t}^*$ is  $p^{m-1}(p-1)-(p-1)p^{t+1}$ or
 $p^{m-1}(p-1)-p^{t+1}$ by Lemma \ref{R001}, Lemma \ref{DH001} and Remark \ref{R0022}.

{\bf Secondly}, consider $d_{\tau_{t-d+4}}$ for the $d$th order where $3\leq d\leq i$. The corresponding cyclic subcode ${C}_{\tau_{t-d+4}}$ obtained from the chain (\ref{CDP002}) has quadratic forms with ranks $r_{t-d+4}\ge 2(t-d)+4$ by Lemma \ref{MS02}. The number of quadratic forms contained in ${C}_{\tau_{t-d+4}}$ is $p^{(d-1)m}$. By Proposition \ref{EF002}, the maximum number of quadratic forms with ranks $r\geq 2(t-d)+7$ is $p^{(m+1)(d-2)}$ which is smaller than $p^{(d-1)m}$.
  So, there must be some quadratic forms of ranks $r_{t-d+4} =2(t-d)+6$, $2(t-d)+5$ or $2(t-d)+4$. Let's consider these three cases separately, which shows the construction of set (\ref{B01}).
\begin{enumerate}
\renewcommand{\labelenumi}{$($\mbox{\roman{enumi}}$)$}
\item\label{CASE001} Assume that there is a quadratic form $F(X)$ of rank $2(t-d)+6$ in ${C}_{\tau_{t-d+4}}$.
\begin{enumerate}
\item From Lemma \ref{FL002}, suppose the corresponding exponential sum is $S=p^{m-{{r_{t-d+4}}\over 2}}=p^{t+d-1}$. By Lemma \ref{ES02}, the sum $R$ corresponding to $S$ is $(p-1)p^{t+d-1}$. According to the relation between the weight of a codeword and corresponding exponential sum (\ref{C01}), there is a codeword of weight $p^{m-1}(p-1)-(p-1)p^{t+d-2}$.

\item Suppose the corresponding exponential sum is $S=-p^{m-{{r_{t-d+3}}\over 2}}=-p^{t+d-1}$. By the second part of Lemma \ref{FL002}, there is an exponential sum with value $S'=-{\zeta_p^{c}}p^{t+d-1}$, where ${c}$ can be chosen to be nonzero by Lemma \ref{LE04}. By Remark \ref{R0022}, the related sum $R'$ has value $p^{t+d-1}$. Hence, there exists a codeword of weight $p^{m-1}(p-1)-p^{t+d-2}$ by equation (\ref{C01}).

 \end{enumerate}
\item Assume that there is a quadratic form $F(X)$ of rank $2(t-d)+5$. By Lemma \ref{FL002}, the corresponding exponential sum has value $S=\varepsilon \sqrt{p^*}p^{t+d-1} (\varepsilon=\pm 1)$, and  $S'=\varepsilon {\zeta_p^{c}} \sqrt{p^*}p^{t+d-1}$. By Remark \ref{R0022}, the related sum $R'$ has value $\varepsilon {\left({{-{c}}\over p}\right)}p^{t+d}$, where $\varepsilon {\left({{-{c}}\over p}\right)}$ is set to be $1$ by Lemma \ref{LE04}. Again from equation (\ref{C01}), there is a codeword of weight $p^{m-1}(p-1)-p^{t+d-1}$.

\item Assume that there is a quadratic form $F(X)$ of rank $2(t-d)+4$. As in case (i), there is a codeword of weight $p^{m-1}(p-1)-(p-1)p^{t+d-1}$ or $p^{m-1}(p-1)-p^{t+d-1}$.

\end{enumerate}
  For any other chain in the first $i$ orders, to get the optimum distance profile, the first selected primitive idempotent is also $\theta_1^*$. For the $d$th order, where $2\leq d\leq i$, denote the generated cyclic subcode by ${C}_{\tau_{t-d+4}}'$. As in the above analysis, the least rank of the quadratic forms contained in ${C}_{\tau_{t-d+4}}'$ is at most $r_{t-d+4} =2(t-d)+6$.
 Then the possible minimum weights are at most $p^{m-1}(p-1)-(p-1)p^{t+d-2}$ or $p^{m-1}(p-1)-p^{t+d-2}$, which are the same as case (i).

 Note that, in last paragraph, one condition of Lemma \ref{LE04} is used such that the rank of corresponding quadratic form is larger than or equal to $2$. If the rank is $1$, use Lemma \ref{R03} instead of Lemma \ref{LE04}. Furthermore, the lower the rank, the  smaller the corresponding codeword weight would be.
 Therefore, for $3\leq d\leq i$ we have $t-i+4\leq t-d+4\leq t+1$, $d_{\tau_{t-d+4}}\in B_{\tau_{t-d+4}}$, and the minimum distance in the ODPC is not bigger than $\max B_{\tau_{t-d+4}}$ of (\ref{B01}).

 {\bf Thirdly}, for the $d$th order, where $i+1\leq d\leq j$ whence $t-j+4\leq t-d+4\leq t-i+3$, the primitive idempotent $\theta_{l_{t+1}}^*$ is added in the beginning of this third part for the construction of the cyclic subcode ${C}_{\tau_{t-d+4}}$. The number of quadratic forms contained in ${C}_{\tau_{t-d+4}}$ is $p^{(d-2)m+{m\over 2}}$. As above, the minimum distance $d_{\tau_{t-d+4}}$ belongs to the set
  $B_{\tau_{t-d+4}}=\{p^{m-1}(p-1)-p^{t+d-2}, p^{m-1}(p-1)-(p-1)p^{t+d-2}\}$.

{\bf Fourthly}, for the $d$th order, where $j+1\leq d\leq t+3$ whence $1\leq t-d+4\leq t-j+3$, the corresponding cyclic subcode is also denoted by ${C}_{\tau_{t-d+4}}$. Since the primitive idempotent $\theta_0$ is accumulated in the beginning of this part,
as above it can be verified that the minimum distance $d_{\tau_{t-d+4}}$ belongs to the set $B_{\tau_{t-d+4}}=\{p^{m-1}(p-1)-1-(p-1)p^{t+d-3}, p^{m-1}(p-1)-1-p^{t+d-3}\}$. In details, choose a nonzero element $b$ of equation (\ref{C02}) corresponding to the case that $\delta_{b,0}-1=-1$. In addition, since the ranks of corresponding quadratic forms are not less than $2$, the element $c$ of Lemma \ref{FL002} can be chosen to be any one of $\mathbb{F}_p$ by Lemma \ref{LE04}. Thus, substituting $c$ of Remark \ref{R0022} with the value $b+c$, the possible values of the minimum distance are in the set $B_{\tau_{t-d+4}}$.

{\bf Finally}, set $B_{\tau_{0}}=\{p^{m-1}(p-2)-1\}$ since the code is $\mathcal{GRM}(2,m)^*$. \qed

\end{proof}

\begin{corollary}\label{CDP001}
 In fact, the distance profile
 \begin{equation}\label{CDP004}
 d_{\tau_0}, d_{\tau_1}, \ldots, d_{\tau_{t+2}}, d_{\tau_{t+3}}
  \end{equation}
  obtained from the chain (\ref{CDP002}) of Theorem \ref{ME002} is a lower bound on the ODPC-I$^{inv}$, and the sequence
 \begin{equation}\label{CDP005}
 \max B_{\tau_0}, \max B_{\tau_1}, \ldots, \max B_{\tau_{t+2}}, \max B_{\tau_{t+3}}
  \end{equation}
  gives an upper bound on ODPC-I$^{inv}$. Since there are only a few elements with small differences in each set $B_{\tau_u}$, from the analysis of Remark \ref{CDP010} we say that in this sense the lower bound (\ref{CDP004}) almost achieves this upper bound.
\end{corollary}

\begin{remark}\label{CDP010}
Divided by $p^{m-1}(p-1)$, the largest difference between the elements of each $B_{\tau_u}$ is about $1\over {p^{m-t-d}}$ where $m=2t+2$.
\end{remark}

The last statement of Theorem \ref{ME002} states that ``the minimum distance is less than or equal to the corresponding $\max B_{\tau_u}$".
The comparison is component by component, which is stronger than the comparison in the inverse dictionary order, where only some components are compared. Note that the sequence $d_{\tau_0}, d_{\tau_1}, \ldots, d_{\tau_{t+2}}, d_{\tau_{t+3}}$ is an upper bound on the sequence $\min B_{\tau_0}, \min B_{\tau_1}, \ldots, $ $\min B_{\tau_{t+2}}, \min B_{\tau_{t+3}}$ in the inverse dictionary order.

The ODPC of $\mathcal{GRM}(2,m)^*$ has been analyzed in the above under Standard I, and the corresponding research in the inverse dictionary order under Standard II will be provided in the following paragraphs. Define the sets:
\begin{itemize}
\renewcommand{\labelitemi}{\labelitemiii}
\item
$B_{\tau_{t+3}}=\{p^m-1\}$; \ $B_{\tau_{t+2}}=\{p^{m-1}(p-1)-1\}$; \ $B_{\tau_{t+1}}=\{p^{m-1}(p-1)-1-p^t\}$;
\item and for $4\leq d\leq t+3$,
\[
  B_{\tau_{t-d+4}}=\{p^{m-1}(p-1)-1-(p-1)p^{t+d-3}, p^{m-1}(p-1)-1-p^{t+d-3}\};
\]
\item finally, $B_{\tau_{0}}=\{p^{m-1}(p-2)-1\}$.
\end{itemize}

\begin{theorem}\label{EP02}
Let $m=2t+2, t\geq 1$. For the cyclic code $\mathcal{GRM}(2,m)^*$,
the chain obtained by accumulating the primitive idempotents
 one-by-one in the following order
\begin{equation}\label{CDP0020}
\theta_0, \theta_1^*, \theta_{l_{t+1}}^*, \cdots, \theta_{l_1}^*, \theta_{l_0}^*
\end{equation}
provides a distance profile satisfying
 \begin{equation}
d_{\tau_u}\in B_{\tau_u} (0\leq u\leq t+3)
\end{equation}
 in the inverse dictionary order under standard II.
 The corresponding cyclic subcode chain (expansion from right to left) is denoted by
\[
 C_{\tau_{0}}\supset C_{\tau_{1}}\supset \cdots \supset C_{\tau_{t+2}}\supset C_{\tau_{t+3}}.
\]
For any other cyclic subcode chain, the minimum distance is less than or equal to the corresponding $\max B_{\tau_u}$, especially the ODPC.
\end{theorem}

\begin{proof}
It can be checked that the first selected primitive idempotent should be $\theta_0$ under Standard II, which corresponds to the cyclic code with minimum distance $d_{\tau_{t+3}}=p^m-1$. By (\ref{C02}) and Remark \ref{R00222}, the second selected primitive idempontent is $\theta_1^*$, and the corresponding minimum distance is $d_{\tau_{t+2}}=p^{m-1}(p-1)-1$. From the third step of (\ref{CDP0020}), refer to the fourth part in the proof of Theorem \ref{ME002}, and then
$d_{\tau_u}\in B_{\tau_u} (0\leq u\leq t+3)$. \qed
\end{proof}

\begin{corollary}\label{CDP02}
 In fact, the distance profile
 \begin{equation}\label{CDP0040}
 d_{\tau_0}, d_{\tau_1}, \ldots, d_{\tau_{t+2}}, d_{\tau_{t+3}}
  \end{equation}
  obtained from the chain (\ref{CDP0020}) of Theorem \ref{EP02} is a lower bound on the ODPC-II$^{inv}$, and the sequence
 \begin{equation}\label{CDP0050}
 \max B_{\tau_0}, \max B_{\tau_1}, \ldots, \max B_{\tau_{t+2}}, \max B_{\tau_{t+3}}
  \end{equation}
  gives an upper bound on ODPC-II$^{inv}$. Since there are only a few elements with small differences in each set $B_{\tau_u}$, from the analysis of Remark \ref{CDP01} we say that in this sense the lower bound (\ref{CDP0040}) almost achieves this upper bound.
\end{corollary}

\begin{remark} \label{CDP01}
In Theorem \ref{EP02} the difference between the values of the set $B_{\tau_{t-d+4}} (4\leq d\leq t+3)$ is $D_{\tau_{t-d+4}}={(p-2) }p^{t+d-3}$. Divide $D_{\tau_{t-d+4}}$ by the smaller one of $B_{\tau_{t-d+4}}$, then
\[
 {{D_{\tau_{t-d+4}}}\over {{p^{m-1}(p-1)-1-{(p-1)}p^{t+d-3}}}}  \approx {1\over {p^{t-d+4}}}\approx 0
\]
for sufficiently large $p$ or $t-d$. An example of the above theorem is presented in Section \ref{Sec4.5} for the case of $m=4$, for which (\ref{CDP0040}) is exactly the ODPC-II$^{inv}$.
\end{remark}

\subsubsection{The case of $m=2t+1$ }\label{Sec4.4.2}

This subsection analyzes the ODPC of $\mathcal{GRM}(2,m)^*$ when $m=2t+1$ in the inverse dictionary order in Proposition \ref{E2002} under Standard I, and in Proposition \ref{E2001} under Standard II respectively. In fact, we give an upper bound and a lower bound on the ODPC, where the lower bound almost achieves the upper bound in some sense. 
Note that some preliminaries of Proposition \ref{E2002} are presented in Lemma \ref{FT001}, Lemma \ref{EC002} and Corollary \ref{CE002}. Comparing with Lemma \ref{C001}, Lemma \ref{FC001} considers the case of odd $m$.

\begin{lemma}\label{FC001}
 Let $m=2t+1, t\geq 2$. For $0\leq i\leq t$, the irreducible cyclic code $\mathcal{C}_i$ with primitive idempotent $\theta_{l_i}^*$ has only one nonzero weight $p^{m-1}(p-1)$.
 \end{lemma}

 \begin{lemma}\label{FT001}
Let $m=2t+1, t\geq 2$. The cyclic code $\mathcal{C}_i' (1\leq i\leq t)$ with idempotent $\theta_1^*+\theta_{l_i}^*$ has only three nonzero weights and
\[
 \begin{array}{ll}
 A_{p^{m-1}(p-1)-p^t}&={1\over 2}(p-1)(p^m-1)(p^{m-1}+p^t),\\
   A_{p^{m-1}(p-1)+p^t}&={1\over 2}(p-1)(p^m-1)(p^{m-1}-p^t),  \\
   A_{p^{m-1}(p-1)}&= (p^m-1)(p^{m-1}+1).
 \end{array}
\]
  The cyclic code $C_0''$ with idempotent $\theta_1^*+\theta_{l_0}^*$ has the same weight distribution.
 \end{lemma}

 \begin{proof}
 By Lemma \ref{GCD002}, the dimension of $\mathcal{C}_i' (1\leq i \leq t)$ is $2m$. From equation (\ref{C01}), Lemma \ref{FL002}, Lemma \ref{DH001} and Remark \ref{R0022}, it can be verified that there are only three possible nonzero weights ${p^{m-1}(p-1)-p^t}, {p^{m-1}(p-1)+p^t}$ and ${p^{m-1}(p-1)}$.

It can be verified that $A_1'$, the number of codewords of weight $1$ in the dual $\mathcal{C}_i'^{\perp}$ of the code $\mathcal{C}_{i}'$, is zero. From Pless first moment identity (\ref{MS003}), the following holds
\begin{eqnarray}
 & & {\left(p^{m-1}(p-1)-p^t\right)}\cdot A_{p^{m-1}(p-1)-p^t}\nonumber \\
 & + & {\left(p^{m-1}(p-1)+p^t\right)}\cdot A_{p^{m-1}(p-1)+p^t} + {p^{m-1}(p-1)}\cdot A_{p^{m-1}(p-1)}\nonumber \\
 & = & {{p-1}\over p}\left(p^m-1\right){p^{2m}}.\label{FT003}
\end{eqnarray}

\noindent From the primitive idempotent $\theta_1^*$ and $\theta_{l_i}^*$, the number of codewords of weight $2$ in $\mathcal{C}_i'^{\perp}$  is the number of nonzero solutions of
 \begin{equation}\label{FT02}
 \begin{cases}
 x+y=0\\
 x^{p^i+1}+y^{p^i+1}=0.
 \end{cases}
 \end{equation}
But $\{(x,y)|x,y\ \mbox{satisfy}\ (\ref{FT02})\}=\{(0,0)\}$, which implies that $A_2'=0$.
From Pless second moment identity, we get
 \begin{equation}\label{FT005}
 \begin{array}{lll}
 & & {\left(p^{m-1}(p-1)-p^t\right)}^2\cdot A_{p^{m-1}(p-1)-p^t} \\
 &+& {\left(p^{m-1}(p-1)+p^t\right)}^2 \cdot A_{p^{m-1}(p-1)+p^t}+\left({p^{m-1}(p-1)}\right)^2 \cdot A_{p^{m-1}(p-1)}\\
 &=& (p^m-1){p^{2m-2}}\left((p-1)^2(p^m-2)+p(p-1)\right).
 \end{array}
 \end{equation}
 The number of nonzero codewords in $\mathcal{C}_i'$ is
 \begin{equation}\label{FT004}
 A_{p^{m-1}(p-1)-p^t}+ A_{p^{m-1}(p-1)+p^t}+A_{p^{m-1}(p-1)}=p^{2m}-1.
 \end{equation}

\noindent Solving equations (\ref{FT003}), (\ref{FT005}) and (\ref{FT004}), the weight distribution can be obtained.\qed
 \end{proof}

Above paragraphs consider the weight distribution of the cyclic code $\mathcal{C}_i'$. Now we focus on the cyclic code ${C}_{i,j}$ with idempotent $\theta_{l_i}^*+\theta_{l_j}^* (0\leq i,j\leq t)$. From equation (\ref{C01}) and Lemma \ref{ES02} its possible weights are of the form
\begin{equation}\label{E1}
p^{m-1}(p-1), p^{m-1}(p-1)\pm (p-1)p^{m-d-1} (1\leq d\leq t)
\end{equation}
whence $t\leq m-d-1\leq 2t-1$.
\begin{lemma}\label{EC002}
Let $m=2t+1, t\geq 2$. For $0\leq i,j\leq t$, the cyclic code  ${C}_{i,j}$ with idempotent $\theta_{l_i}^*+\theta_{l_j}^*$ has weights not only of the form
$p^{m-1}(p-1)$, $p^{m-1}(p-1)+(p-1)p^{m-d-1} (1\leq d\leq t)$ in (\ref{E1}).
\end{lemma}

\begin{proof}
From Lemma \ref{GCD002}, ${C}_{i,j}$ has dimension $2m$ and length $l=p^m-1$.
Assume that the weights are only of the form given in the lemma. Let $n_0$ be the number of codewords in ${C}_{i,j}$ of weight $p^{m-1}(p-1)$, and $n_d$ be the number of codewords of weight $p^{m-1}(p-1)+(p-1)p^{m-d-1}$ where $1\leq d\leq t$.

From Pless first moment identity (\ref{MS003})
\begin{equation}\label{MS0030}
p^{m-1}(p-1)n_0 + \sum_{d=1}^{t}\left(p^{m-1}(p-1)+(p-1)p^{2t-d}\right)n_{d} ={{p-1}\over p}(p^m-1)p^{2m},
\end{equation}
where the left part is
\[
p^{m-1}(p-1)(n_0+n_t+\cdots+n_1) + (p-1)\left(p^{t} n_t+\cdots+ p^{2t-1} n_1\right)
\]
and $n_0+n_t+\cdots+n_1=p^{2m}-1$.
After simplification, (\ref{MS0030}) becomes
\[
(p-1)(p^m-1)p^{m-1}+(p-1)\left(p^{t} n_t+\cdots+ p^{2t-1} n_1\right)=0
\]
which is impossible.     \qed
\end{proof}

A combination of Lemma \ref{FT001} and Lemma \ref{EC002} is Corollary \ref{CE002}.
\begin{corollary}\label{CE002}
Let $m=2t+1, t\geq 2$. The minimum distance of the cyclic code ${C}_{i,j}$ with idempotent $\theta_{l_i}^*+\theta_{l_j}^*$ is smaller than the minimum distance of the cyclic code $\mathcal{C}_k' (\mathcal{C}_0'')$ with idempotent $\theta_1^*+\theta_{l_k}^* (\theta_1^*+\theta_{l_0}^*)$.
Here, $0\leq i,j\leq t$ and $1\leq k\leq t$.
\end{corollary}

 For $m=2t+1$, the following facts are about the cyclic code $\mathcal{GRM}(2,m)^*$ by Lemma \ref{th1}. The length of the cyclic subcode chains of $\mathcal{GRM}(2,m)^*$ is $\lambda =t+3$, the total number of chains is $\lambda !=(t+3)!$, the number of chains contained in each class is $\mu=(t+2)!1!$ and the number of classes is ${\lambda ! \over {\mu}}=t+3$.

For the cyclic code $\mathcal{GRM}(2,m)^*$ with idempotent $\theta_0+\theta_1^*+\theta_{l_t}^*+\cdots+\theta_{l_1}^*+\theta_{l_0}^*$, let's consider the following dimension profile under Standard I
\begin{equation}\label{CE0022}
(t+2)m+1, \ldots, (i+1)m+1, im+1, im, \ldots, 2m, m
\end{equation}
where $2\leq i\leq t$,
  which is of universal significance since it includes almost every dimension profile.
The following $t+3$ sets restrict the range of the distance profile in Proposition \ref{E2002}.
\begin{itemize}
\renewcommand{\labelitemi}{\labelitemiii}
\item First part: $ B_{\tau_{t+2}}=\{p^{m-1}(p-1)\}; \ B_{\tau_{t+1}}=\{p^{m-1}(p-1)-p^t\}$;
 \item Second part: for $3\leq d\leq i$,
\[
\begin{array}{ll}
B_{\tau_{t-d+3}}=&\{p^{m-1}(p-1)-(p-1)p^{t+d-3}, p^{m-1}(p-1)-p^{t+d-3},\\
                 & p^{m-1}(p-1)-p^{t+d-2}\};
\end{array}
\]
\item  Third part: for $i+1\leq d\leq t+2$,
\[
\begin{array}{ll}
B_{\tau_{t-d+3}}=&\{p^{m-1}(p-1)-1-p^{t+d-4}, p^{m-1}(p-1)-1-(p-1)p^{t+d-4},\\
                                  &p^{m-1}(p-1)-1-p^{t+d-3}\};
\end{array}
\]
\item Final part: $B_{\tau_{0}}= \{p^{m-1}(p-2)-1\}$.
\end{itemize}

\begin{proposition}\label{E2002}
Let $m=2t+1, t\geq 2$. For the cyclic code $\mathcal{GRM}(2,m)^*$,
consider the dimension profile (\ref{CE0022}) under standard I.
 The chain obtained by accumulating the primitive idempotents
 one-by-one in the following order
\begin{equation}\label{CPF001}
\theta_{1}^*, \theta_{l_0}^*, \theta_{l_1}^*, \ldots, \theta_{l_{i-2}}^*, \theta_0, \theta_{l_{i-1}}^*, \ldots, \theta_{l_t}^*
\end{equation}
presents a distance profile satisfying
\begin{equation}
 d_{\tau_u}\in B_{\tau_u} (0\leq u\leq t+2)
 \end{equation}
  in the inverse dictionary order, where $2\leq i \leq t$.
 The corresponding cyclic subcode chain (expansion from right to left) is denoted by
\[
 C_{\tau_{0}}\supset C_{\tau_{1}}\supset \cdots \supset C_{\tau_{t+1}}\supset C_{\tau_{t+2}}.
\]
For any other cyclic subcode chain, the minimum distance is less than or equal to the corresponding $\max B_{\tau_u}$, especially the ODPC.
\end{proposition}

\begin{proof}
The following proof is similar to Theorem \ref{ME002} and only the main ideas are kept.
According to the assumption of the dimension profile (\ref{CE0022}), and the sizes of the cyclotomic cosets in Lemma \ref{GCD002}, $\theta_0$ is selected in the $(i+1)$th order in (\ref{CPF001}) where $2\leq i\leq t$.

{\bf Firstly}, for the first order, by Lemma \ref{FC001}, Lemma \ref{SC01} and Corollary \ref{CE002}, to get the optimum minimum distance, the first selected primitive idempotent can be $\theta_1^*$
with corresponding minimum distance $p^{m-1}(p-1)\in B_{\tau_{t+2}}$. For the second order, applying Lemma \ref{FT001}, the minimum distance of the cyclic subcode ${C}_{\tau_{t+1}}$ is $p^{m-1}(p-1)-p^t \in B_{\tau_{t+1}}$.

{\bf Secondly}, let's consider the $d$th order where $3\leq d\leq i$. Denote the cyclic subcode of the chain obtained from equation (\ref{CPF001}) by ${C}_{\tau_{t-d+3}}$. Its minimum distance is analyzed in cases (i) and (ii).

By Lemma \ref{RQ002}, the quadratic forms contained in ${C}_{\tau_{t-d+3}}$ have ranks $r_{\tau_{t-d+3}}\geq m-2(d-2)$. By Lemma \ref{GCD002}, the number of quadratic forms contained in ${C}_{\tau_{t-d+3}}$ is $p^{(d-1)m}$. By Proposition \ref{EF002}, the maximum number of quadratic forms of ranks not less than $m-2(d-2)=2(t-d+3)-1$ is also $p^{(d-1)m}$ where $2(t-d+3)-1$ or $2(t-d+3) $ is achieved, which is the least rank of quadratic forms contained in ${C}_{\tau_{t-d+3}}$.
\begin{enumerate}
\renewcommand{\labelenumi}{$($\mbox{\roman{enumi}}$)$}
\item Suppose there is a quadratic form $F(X)$ of rank $2(t-d+3)$. As in the proof of Theorem \ref{ME002}, there is a weight of $p^{m-1}(p-1)-(p-1)p^{t+d-3}$ or $p^{m-1}(p-1)-p^{t+d-3}$.
\item Suppose there is a quadratic form $F(X)$ of rank $2(t-d+3)-1$. Similarly, there is a weight of $p^{m-1}(p-1)-p^{t+d-2}$.

\end{enumerate}
Thus, the minimum distance $d_{\tau_{t-d+3}}$ belongs to the sets $B_{\tau_{t-d+3}}=\{p^{m-1}(p-1)-(p-1)p^{t+d-3}, p^{m-1}(p-1)-p^{t+d-3},
 p^{m-1}(p-1)-p^{t+d-2}\}$ for $3\leq d\leq i$ whence $t-i+3\leq t-d+3\leq t$.

{\bf Thirdly}, for the $d$th order, where $i+1\leq d\leq t+2$ whence $1\leq t-d+3\leq t-i+2$, the primitive idempotent $\theta_0$ is accumulated in the beginning of this part.
And the minimum distance $d_{\tau_{t-d+3}}$ belongs to the set $B_{\tau_{t-d+3}}$.

{\bf Finally}, set $B_{\tau_{0}}=\{p^{m-1}(p-2)-1\}$ which corresponds to the minimum distance of $\mathcal{GRM}(2,m)^*$.  \qed
\end{proof}

As a counterpart to Theorem \ref{EP02}, the following result is for the case of $m=2t+1$, which tries to bound the ODPC-II$^{inv}$ tightly.
Define those sets
\begin{itemize}
\renewcommand{\labelitemi}{\labelitemiii}
\item
$B_{\tau_{t+2}}= \{p^m-1\}$; \ $B_{\tau_{t+1}}= \{p^{m-1}(p-1)-1\}$; \ $B_{\tau_{t }}= \{p^{m-1}(p-1)-1-p^t\}$;
\item and for $4\leq d\leq t+2$,
\[
\begin{array}{ll}
B_{\tau_{t-d+3}}=&\{p^{m-1}(p-1)-1-p^{t+d-4}, p^{m-1}(p-1)-1-(p-1)p^{t+d-4},\\
                 &                 p^{m-1}(p-1)-1-p^{t+d-3}\};
\end{array}
\]
\item finally, $B_{\tau_{0}}= \{p^{m-1}(p-2)-1\}.$
\end{itemize}

\begin{proposition}\label{E2001}
Let $m=2t+1, t\geq 2$. For the cyclic code $\mathcal{GRM}(2,m)^*$,
the chain obtained by accumulating the primitive idempotents
 one-by-one in the following order
\begin{equation}\label{CDP0001}
\theta_0, \theta_1^*, \theta_{l_0}^*, \theta_{l_1}^*, \cdots, \theta_{l_{t-1}}^*, \theta_{l_t}^*
\end{equation}
 presents a distance profile satisfying
 \begin{equation}
 d_{\tau_u}\in B_{\tau_u} (0\leq u\leq t+2)
 \end{equation}
  in the inverse dictionary order under Standard II.
 The corresponding cyclic subcode chain (expansion from right to left) is denoted by
\[
 C_{\tau_{0}}\supset C_{\tau_{1}}\supset \cdots \supset C_{\tau_{t+1}}\supset C_{\tau_{t+2}}.
\]
For any other cyclic subcode chain, the minimum distance is less than or equal to the corresponding $\max B_{\tau_u}$, especially the ODPC.
\end{proposition}

Like Corollary \ref{CDP001} and Corollary \ref{CDP02}, the distance profile
$d_{\tau_0}, d_{\tau_1}, \ldots, d_{\tau_{t+1}}, d_{\tau_{t+2}}$ in Proposition \ref{E2002} and Proposition \ref{E2001} is a lower bound on the ODPC-II$^{inv}$, and the sequence
$\max B_{\tau_0}, \max B_{\tau_1}, \ldots, \max B_{\tau_{t+1}}, \max B_{\tau_{t+2}} $
  gives an upper bound on ODPC-II$^{inv}$. Since there are only a few elements with small differences in each set $B_{\tau_u}$ (relatively), we say that in this sense the lower bound almost achieves this upper bound. 

\subsection{Examples}\label{Sec4.5}
In this subsection, two examples with construction are presented using Matlab R2010 for Theorem \ref{EP02} and Proposition \ref{E2001} respectively, where the distance profiles can be verified to be the ODPCs of $\mathcal{GRM}(2,m)^*$ in the inverse dictionary order.
\begin{example}\label{EEM01}
According to Theorem \ref{EP02}, for the case of $m=2\cdot 1+2=4$ and $p=3$, there are only five primitive idempotents
\begin{equation} \label{EM001}
\theta_0,\theta_1^*,\theta_{l_2}^*,\theta_{l_1}^*,\theta_{l_0}^*
\end{equation}
in the construction of cyclic subcode chains.
\begin{itemize}
\renewcommand{\labelitemi}{\labelitemiii}
\item The first step.
The above five primitive idempotents correspond to five irreducible cyclic subcodes with minimum distances $80,54,60,48,48$ respectively. So, the first selected primitive idempotent is $\theta_0$.

\item The second step.
The idempotents $\theta_0+\theta_1^*$, $\theta_0+\theta_{l_2}^*$,
$\theta_0+\theta_{l_1}^*$ and  $\theta_0+\theta_{l_0}^*$
correspond to four cyclic subcodes with minimum distances $53,50,44$ and $48$ respectively. So, the idempotent
$\theta_0+\theta_1^*$  is selected for this step.

\item The third step.
The idempotents $\theta_0+\theta_1^*+\theta_{l_2}^*$,
 $\theta_0+\theta_1^*+\theta_{l_1}^*$ and
$\theta_0+\theta_1^*+\theta_{l_0}^*$
correspond to minimum distances $50,44$ and $47$ respectively. So, the idempotent
$\theta_0+\theta_1^*+\theta_{l_2}^*$  is selected for this step.

\item The  fourth step.
The idempotents $\theta_0+\theta_1^*+\theta_{l_2}^*+\theta_{l_1}^*$ and
$\theta_0+\theta_1^*+\theta_{l_2}^*+\theta_{l_0}^*$
correspond to minimum distances $ 44$ and $35$ respectively. So, the idempotent
$\theta_0+\theta_1^*+\theta_{l_2}^*+\theta_{l_1}^*$  is selected for this step.

\item Finally, the idempotent
$\theta_0+\theta_1^*+\theta_{l_2}^*+\theta_{l_1}^*+\theta_{l_0}^*$ corresponds to $\mathcal{GRM}(2,4)^*$ with minimum distance $26$.
\end{itemize}
 Therefore, the corresponding distance profile in the inverse dictionary order $26,44,50,53,80$ is the ODPC-II$^{inv}$ of $\mathcal{GRM}(2,4)^*$,
and the construction of an optimum cyclic subcode chain can be obtained by accumulating the primitive idempotents in the order of (\ref{EM001}).
 \end{example}

\begin{example}\label{EEM02}
According to Proposition \ref{E2001}, for the case of $m=2\cdot 2+1=5$ and $p=3$, there are also only five primitive idempotents
\begin{equation}\label{EEM002}
\theta_0,\theta_1^*,\theta_{l_0}^*,\theta_{l_1}^*,\theta_{l_2}^*,
\end{equation}
in the construction of cyclic subcode chains.
\begin{itemize}
\renewcommand{\labelitemi}{\labelitemiii}
\item The first step.
The above five primitive idempotents correspond to five irreducible cyclic subcodes with minimum distances $242,162,162,162,162$ respectively. So, the first selected primitive idempotent is $\theta_0$.

\item The second step.
The idempotents $\theta_0+\theta_1^*$, $\theta_0+\theta_{l_2}^*$,
$\theta_0+\theta_{l_1}^*$ and  $\theta_0+\theta_{l_0}^*$
correspond to four cyclic subcodes with minimum distances $161,152,152,152$ respectively. So, the idempotent
$\theta_0+\theta_1^*$  is selected for this step.

\item The third step.
The idempotents $\theta_0+\theta_1^*+\theta_{l_2}^*$,
 $\theta_0+\theta_1^*+\theta_{l_1}^*$ and
$\theta_0+\theta_1^*+ \theta_{l_0}^*$
all correspond to codes with minimum distance $152$.
So, any one of the three combinations can be selected for this step.

\item The fourth step depends on the third step. Considering all cases of the third step,
since the idempotents $\theta_0+\theta_1^*+\theta_{l_2}^*+\theta_{l_1}^*$,
 $\theta_0+\theta_1^*+\theta_{l_2}^*+\theta_{l_0}^*$ and
$\theta_0+\theta_1^*+\theta_{l_1}^*+\theta_{l_0}^*$
all correspond to codes with minimum distance $134$, any one of the three combinations can be selected for this step.

\item Finally, the idempotent
$\theta_0+\theta_1^*+\theta_{l_2}^*+\theta_{l_1}^*+\theta_{l_0}^*$ corresponds to $\mathcal{GRM}(2,5)^*$ with minimum distance $80$.
\end{itemize}
Therefore, the corresponding distance profile $80,134,152,161,242$ is the ODPC-II$^{inv}$ of $\mathcal{GRM}(2,5)^*$,
and the construction of an optimum cyclic subcode chain can be obtained by accumulating the primitive idempotents in the order of (\ref{EEM002}).
 \end{example}

In Example \ref{EEM01}, there is only one possible selection process to achieve the ODPC, while in Example \ref{EEM02}, there are many selection processes.

\section{Conclusions}\label{Sec5}

Comparing to the optimum distance profiles of linear block codes \cite{CH001,HL001,HL002}, the same problems on cyclic codes are more popular because of faster encoding and decoding.
In fact, we use two standards to expand cyclic subcodes to $\mathcal{GRM}(2,m)^*$, where Theorem \ref{ME002} and Proposition \ref{E2002} are for Standard I, Theorem \ref{EP02} and Proposition \ref{E2001} are for Standard II, Theorem \ref{ME002} and Theorem \ref{EP02} are for even $m$, Proposition \ref{E2002} and  Proposition \ref{E2001} are for odd $m$. For each standard, the provided lower bounds on the optimum distance profiles almost achieve the upper bounds in some sense. Sometimes we even get the optimum one exactly, as illustrated in Theorem \ref{F01} and Section \ref{Sec4.5}.


\end{document}